\documentclass[12pt,a4paper]{article} \parindent=1cm
\usepackage[left=1cm,right=1cm,top=1cm,bottom=1cm]{geometry}
\usepackage{amsmath}
\usepackage{amssymb}
\usepackage{graphicx}

\newcommand{\RN}[1]{%
  \textup{\uppercase\expandafter{\romannumeral#1}}%
}
\DeclareMathOperator{\arcth}{arcth}
\DeclareMathOperator{\pd}{\partial}

\DeclareMathOperator{\dg}{\sqrt{\mathnormal{ - g}}}
\DeclareMathOperator{\ia}{\frac{1}{\mathnormal{a}}}
\DeclareMathOperator{\wt}{\omega \tau}
\DeclareMathOperator{\mk}{|\mathbf k|}
\DeclareMathOperator{\sh}{sh}
\DeclareMathOperator{\ch}{ch}
\DeclareMathOperator{\thg}{th}
\DeclareMathOperator{\cth}{cth}
\DeclareMathOperator{\arctg}{arctg}

\begin{document}

{
\thispagestyle{empty}
\newpage
\begin{center}
\bf{\Large{On radiation due to homogeneously accelerating sources}}
\vspace{15 pt}

\bf{D.Kalinov}

\vspace{5 pt}

\it{Higher School of Economics, Departament of Mathematics, Moscow, Russia}
\end{center}

\vspace{5 pt}
\begin{center}
\bf{{Abstract}}
\end{center}
The core of this work is an old and broadly discussed problem of the electromagnetic radiation in the case of the hyperbolic motion. We prove that the radiation is non-zero in the lab (Minkowski) frame. Further, we attempt to understand this subject better by using co-moving non-inertial frames of reference, investigating other types of uniformly accelerated motion and, finally, using scalar waves instead of point-like particles as sources of radiation. \
	
\begin{center}
\section*{Contents}
\end{center}

\textbf{
1. Introduction
}

\textbf{
2. Uniformly accelerated particle
}

\ \ \ \ \ 2.1 Introducing two coordinate systems

\ \ \ \ \ 2.2 Introducing the action and the equations of motion

\ \ \ \ \ 2.3 Expressing vector field solutions and the energy flow

\ \ \ \ \ 2.4 Discussion

\textbf{
3. Uniformly rotating particle
}

\ \ \ \ \ 3.1 Introducing the coordinate frames

\ \ \ \ \ 3.2 Expressing vector and electromagnetic field solutions

\ \ \ \ \ 3.3 Asymptomatic behaviour of the electromagnetic field in the co-moving frame

\ \ \ \ \ 3.4 Discussion

\textbf{
4. Wave as a source of radiation
}

\ \ \ \ \ 4.1 Introducing the action

\ \ \ \ \ 4.2 Free moving source in the Minkowski space-time

\ \ \ \ \ 4.3 Free moving source in the co-moving Rindler frame

\ \ \ \ \ 4.4 Uniformly accelerated source in the Minkowski space-time

\ \ \ \ \ 4.5 Discussion

\textbf{
5. Acknowledgements
}	

\textbf{
6. References
}	
	
\clearpage
}
\section{Introduction}
	
So far  many authors have considered the problem of electromagnetic radiation produced by a uniformly accelerated particle in their works (\cite{Born}, \cite{Pauli}, \cite{Fulton}, \cite{Paelris}, \cite{Feynman}, \cite{Bondi}, \cite{Boulware:1979qj}, \cite{Parrott}, \cite{Lyle}, \cite{newref1}, \cite{newref2}). Several of them reasoned that there is no radiation in this case, the arguments given included the absence of magnetic field in the inertial co-moving frame of reference and the fact that the radiation friction force is zero for the hyperbolic motion. Here we study this subject from a different perspective and explain in what sense the homogeneously accelerated particle does radiate.

To begin with, let us discuss how we understand the notion of radiation. We say that the radiation is present in a system if there is a non-zero energy flow through an infinitely distant surface. One can notice that, first, such an energy flow can be zero with the Poynting vector still being non-zero and, second, the energy flow is not invariant under general space-time coordinate transformations. There is another quantity associated with the radiation, namely, the energy loss of the particle. This quantity is invariant under covariant transformations and is equal to $\frac{2}{3}e w^\mu w_\mu$ for a point-like source with the four-acceleration $w_\mu$.  Therefore, it is one of  our objectives -- to investigate the behaviour of the energy flow in different coordinate systems for a uniformly accelerated particle. Also we will discuss different uniform motions and different sources of radiation.

The paper consists of three parts. In the first part we find the solution of Maxwell equations for two reference systems: the inertial lab frame (Minkowski frame), where the world line of the particle is represented by a hyperbola, and the non-inertial co-moving frame\footnotemark[1] (Rindler frame) where the world line is a straight line. From the explicit form of the solutions one can conclude that there is only a static electric field in the Rindler frame, while the radiation is undoubtedly present in the Minkowski frame.\footnotemark[2]

\footnotetext[1]{It should be mentioned that we define the co-moving frame of reference as a static coordinate system in which the particle under consideration is at rest. Generaly, this frame is not related to the Minkowski space-time by a Lorentz boost, hence it is non-inertial.}
\footnotetext[2]{The physical meaning of the obtained results is explained in the ``discussion'' subsection of each section.}

In the second part the same problem is investigated, though this time we consider a uniformly rotating particle, again in two systems: the lab frame and the non-inertial co-moving frame. In this case the solutions are much more complex and implicit. It follows that the radiation is present in the Minkowski space-time but cannot be defined in the co-moving frame of reference because the latter is not extendable beyond some distance away from the source of radiation, i.e. the resulting coordinate patch is compact. 

The last part is devoted to the interpretation of the radiation source as an excitation of a complex scalar field. We  examine the theory of one scalar and one vector gauge field with a coupling constant being turned on adiabatically in the distant past and turned off in the distant future. In this theory a question of interest is as follows: could an excitation of the scalar field at the past infinity create vector harmonics at the future infinity? This, if it is possible, can be interpreted as a radiation of the vector field by the scalar field source. Here we prove that this process is forbidden in  the flat space-time for the free scalar waves as sources of radiation. Further, we show that it is allowed for the uniformly accelerated scalar excitations in the Minkowski space-time and the free ones in the Rindler frame. Moreover, these considerations will allow us to clarify the situation with the point-like source in the Rindler frame. This is the most essential part of this work.

The problem discussed in the first part has been studied by various authors, so the results in this segment of the article are similar to the ones already obtained by other scientists. We will give a quick review of the existing literature in the beginning of the corresponding section.However, in the second part the field calculation in the co-moving non-inertial frame is a new result. The results of the third part are new as well due to the methods which have not been used for this problem previously. 

\section{Uniformly accelerated particle}

This question was originally posed by Born in \cite{Born} where he calculated the fields for the uniformly accelerated particle in the Minkowski space using the  Liénard-Wiechert potentials. Later Pauli in \cite{Pauli} argued that there is no radiation due to the vanishing of the magnetic field at $t = 0$. In \cite{Feynman} Feynman claimed that the radiation power should be proportional to $\dot x \dddot x$, thus for the uniformly accelerated charge it is zero and it does not radiate. In another paper Bondi and Gold \cite{Bondi} calculated the fields using the representation of the uniform accelerated motion as the limit of motions with acceleration switched on for a finite period of time. Using this formalism they concluded that the radiation is present in the system. In another article by Boulware (\cite{Boulware:1979qj})  the non-inertial co-moving frame is used, the stress-energy tensor and the energy flux are calculated and the presence of radiation in the Minkowski space is established.   Moreover, in \cite{Parrott2} Parrott carefully discusses the Lorentz-Dirac equation, and Lyle in \cite{Lyle} uses this discussion to prove that the radiation friction force is zero for this case.   The most complete account of the history of this problem can be found in the book by Lyle \cite{Lyle}. Anyone interested in a more broad discussion of  previously obtained results in this field should consult it. Also similar considerations have been made in the de Sitter space in \cite{Akhmedov:2010ah}. 

In the beginning, we introduce two coordinate systems which are used in this part of our work. Then the expression for the action is written and the equations of motion resulting from it are discussed in both systems. After that we implicitly express the static solution for the vector field in the Minkowski system and make a coordinate change to the Rindler one. Finally, we calculate the energy flow for both systems and conclude that there is no radiation in the Rindler frame, but it is present in the Minkowski one. At the end of this part, we analyse the physical meaning of the obtained results in the ``discussion'' subsection.
\subsection{Introducing two coordinate systems}
To begin with, let us introduce the lab frame and the non-inertial co-moving frame.

The metric in the Minkowski space-time is given by: $ds^2 = dt^2 - dx^2 -dy^2 - dz^2$, and in the Rindler one: \\ $ds^2 = \rho^2 d\tau^2 - d\rho^2 - dy^2 -dz^2$ with $\rho > 0$. 
One can go from the first frame to the second one  by the following coordinate transformation:
\begin{equation}
\tau  = \arcth\frac{t}{x}, \ \rho = \sqrt{x^2 - t^2}, \ y = y , \ z =z \ ,
\end{equation}
and back by:
\begin{equation}
t = \rho \sh\tau ,\  x = \rho\ch\tau, \  y =y,  \ z=z \ .
\end{equation}
It follows that the Rindler coordinates cover only a part of the flat space-time  described by  $x >  |t| $. Besides,  it can be easily observed that they describe the non-inertial co-moving frame for a uniformly accelerated particle. Indeed, if we consider a world line $ t = \frac{1}{a} \sh \theta , \ x = \frac{1}{a} \ch \theta, \ y = 0, \ z = 0$, which corresponds to a particle moving with constant four-acceleration $a_{\mu}a^{\mu} = -a^2$, we can rewrite its world line in the Rindler's coordinates in the following form $\tau = \theta,  \ \rho = \frac{1}{a}, \ y = 0, \ z = 0$, i.e. the particle is at rest, and the frame is co-moving.
\subsection{Introducing the action and the equations of motion}
Now we shall introduce the Maxwell equations for the case under consideration. The action for the vector field $A_\mu$ in a curved space-time is given by:
\begin{equation}
S = \int d^4x \sqrt{-g} \left[ -\frac{1}{16 \pi} F^{\mu \nu} F_{\mu \nu}  - A_{\mu}j^{\mu}\right] \ .
\end{equation}
Here $g = \det g_{\mu \nu}$ and $F_{\mu \nu} = D_{\mu}A_{\nu} - D_{\nu}A_{\mu}= \pd_{\mu}A_{\nu} - \pd_{\nu}A_{\mu}$. Varying the action with respect to $A_\mu$ we obtain the following equations of motion:
\begin{equation}
\frac{1}{\dg} \pd_{\nu} ( \dg F^{\nu \mu} ) = 4 \pi j^{\mu} \ .
\end{equation}
In order to solve them the Lorentz Gauge is used:
\begin{equation}
D^{\mu}A_{\mu} = 0 \ \ \Longleftrightarrow \ \ \pd^{\mu} A_{\mu} = g^{\mu \nu} \Gamma^\eta_{\mu \nu} A_{\eta} \ .
\end{equation}
The equations above can be simplified in both reference systems with the use of the explicit form for the  metric tensor and the gauge condition. For the Minkowski space one easily gets:
\begin{equation}
\Box A_{\mu} = 4 \pi j_{\mu} \ ,
\end{equation}
with $\Box = \pd_{\mu}\pd^{\mu}$. In the Rindler frame it follows that $\dg = \rho$ and all non-zero Christoffel symbols are $\Gamma^1_{00} = \rho, \ \Gamma^0_{01} = \frac{1}{\rho}$, so the gauge condition can be rewritten as:
\begin{equation}
g^{\mu \nu} \pd_{\mu}A_{\nu} = \frac{A_1}{\rho} \ .
\end{equation}
Inserting the definition of the Rindler metric and (7) into (4) one can obtain the following form for the equations of motion:
\begin{equation}
\begin{cases}
\left(\Box_R + \frac{2\pd_1}{\rho} \right) - \frac{2\pd_0}{\rho} = 4 \pi j_0 \\
\left(\Box_R + \frac{1}{\rho^2} \right) A_1 = 4 \pi j_1 \\
\Box_R A_2 = 4 \pi j_2 \\
\Box_R A_3 = 4 \pi j_3  \ .
\end{cases}
\end{equation}
Here: $ \Box_R = g^{\mu \nu} \pd_{\mu}\pd_{\nu} - \frac{\pd_1}{\rho} = \frac{\pd_0\pd_0}{\rho^2} - \frac{\pd_1}{\rho} - \pd_1\pd_1 - \pd_2\pd_2 - \pd_3\pd_3$.
\subsection{Expressing vector field solutions and the energy flow}
In this section we find the solutions for (6) and (8) with a current $j_\mu$ created by a uniformly accelerated particle and, afterwards, calculate the energy flow for both systems. 

To begin with, let us obtain the solutions of (6) by using the Liénard-Wiechert potentials for the fields in the Minkowski space-time. In order to do so the world line should be parametrized. Here the parametrization by the proper time $\theta$ is used:

\begin{equation}
z^1 = \ia \ch (a \theta)  \ \ \   z^0 = \ia \sh (a \theta )\ \ \  z^2 = 0 \ \ \ z^3 = 0 \ .
\end{equation}
Let $(t,x,y,z)$ - be the point in the Minkowski frame where the field is measured. Then, if one introduces the three-velocity vector as $\mathbf v(\theta) = \frac{dz^i}{dz^0} = ( \thg (a\theta), 0, 0)$ and the three-vector $\mathbf R^i(\theta, x^{\mu}) =  x^i - z^i = (x - \ia \ch (a\theta), y, z)$,  one can write down the expression for $A_\mu$ in the following form:
\begin{equation}
\begin{cases}
A_0(x^{\mu}) = \frac{1}{|\mathbf{R}| - (\mathbf{v}, \mathbf{R})} \\
A_i (x^{\mu})= \frac{-v^i}{|\mathbf{R}| - (\mathbf{v}, \mathbf{R})}  \\
|R| = t - z^0 = t - \ia \sh(a\theta) \ ,
\end{cases}
\end{equation}
here the last equation is used to express the radiation proper time moment $\theta$ in terms of $x^{\mu}$. The latter can be rewritten as $t = |\mathbf R| + \ia \sh (a \theta)$, and used to simplify the expressions for the field: $|\mathbf R| -(\mathbf v,\mathbf R) = |\mathbf R| +\ia \sh(a \theta) - x \thg(a \theta) = t - x \thg (a\theta) $. After we substitute this into (10) it transforms as follows:
\begin{equation}
\begin{cases}
A_0  = \frac{1}{t - x \thg (a\theta)}\\
A_1 = \frac{-  \thg(a \theta)}{t - x \thg (a\theta)}\\
A_2 = A_3 = 0 \\
t = |\mathbf{R}| + \ia \sh (a \theta) \ .
\end{cases}
\end{equation}
This is a suitable moment to make a coordinate transformation to the Rindler frame, i.e.  $(A_0, A_1) \to (A^R_0, A^R_1)$. From now on the components of $A_{\mu}$, which are equal to zero will be omitted. Rewriting $A_{\mu}$ as functions of $\rho $ and $\tau$, we get:
\begin{align}
A_\mu = \left(\frac{ \ch (a\theta)}{\rho (\sh \tau \ch (a\theta)- \ch \tau \sh (a\theta))}\ ,\frac{-  \sh (a\theta)}{\rho (\sh \tau \ch (a\theta) - \ch \tau \sh (a\theta))} \right) = \nonumber \\ = \left(\frac{ \ch (a\theta)}{\rho \sh (\tau - a\theta )}\ ,\frac{-  \sh (a\theta)}{\rho\sh (\tau - a\theta )} \right) \ .
\end{align} 
Next, to obtain $A^R_{\mu}$, we should use the Jacobi matrix $J^{\nu}_{\mu} = \frac{\pd x^{\nu}}{\pd x'^{\mu}}$, which in the case under consideration is given by:
\begin{equation}
J^{\nu}_{\mu} = \begin{pmatrix} \rho \ch \tau & \rho \sh \tau \\  \sh \tau & \ch \tau \end{pmatrix} \ ,
\end{equation} 
hence:
\begin{align}
A^R_{\mu} = J^{\nu}_{\mu} A_{\nu}= \frac{1}{\rho \sh (\tau - a\theta )} \Big(\rho(\ch (a\theta) \ch \tau - \sh (a\theta) \sh \tau) , (\sh \tau \ch (a\theta) - \ch \tau \sh (a\theta)) \Big) = \nonumber \\
 = \left(  \cth (\tau - a\theta) , \frac{1}{\rho} \right) \ .
\end{align} 
After fixing the Lorentz gauge there remains a certain gauge invariance, more precisely, one can make the transformation $A^R_{\mu} \to A^R_{\mu} - \pd_{\mu} \alpha$ if $\alpha$ satisfies the equation:
\begin{equation}
g^{\mu \nu} \pd_{\mu}\pd_{\nu} \alpha = \frac{\pd_1\alpha}{\rho} \ .
\end{equation}
It can be easily seen that $\alpha = \ln(\rho)$ is a correct choice. After the transformation by such a gauge parameter one obtains:
\begin{equation}
A^R_{\mu} =  (  \cth (\tau - a\theta) ,0) \ .
\end{equation}
From this we conclude that a homogeneously accelerating particle does not create a magnetic field in the non-inertial co-moving reference frame, because $A_i$ are all zeros. 

At last, one can use the fourth equation in (11) to express the vector field solely in the terms of $\rho, \tau, y, z$:
\begin{align}
t = |\mathbf R| + \ia \sh (a \theta) \implies \nonumber \\ \implies
x^2 + \left(\ia\right)^2 \ch^2 (a\theta) - 2 x\ia \ch (a\theta)  + y^2 + z^2 = t^2 +\left(\ia\right)^2 \sh^2 (a\theta) - 2t\ia \sh (a\theta) \implies \nonumber \\ \implies  a^2\rho^2 + a^2y^2 + a^2z^2 +1 = 2a \rho( \ch \tau \ch(a \theta) - \sh \tau \sh (a\theta)) \implies \nonumber\\
\implies \ch (\tau -a \theta) = \frac{a^2\rho^2 +a^2 y^2 + a^2z^2 + 1}{2a \rho} \ , 
\end{align}
and:
\begin{equation}
A^R_0=\frac{a^2\rho^2 + a^2y^2 + a^2z^2 + 1}{\sqrt{(a^2\rho^2 + a^2y^2 + a^2z^2 + 1)^2 - 4 a^2 \rho^2}} = \frac{\rho^2 + y^2 + z^2 + a^{-2}}{\sqrt{(\rho^2 + y^2 + z^2 + a^{-2})^2 - 4 a^{-2} \rho^2}} \ .
\end{equation}
Thus, the field in Rindler coordinates is time independent. This fact suggests that there is no radiation in the frame under consideration. We prove it by showing that the Poynting vector is zero for this frame later on. 

Now one can transform back to the Minkowski space-time with the use of the inverse Jacobi matrix given by\footnote[3]{However, the resulting expressions for the fields are applicable only in the $x+t > 0$ region of the Minkowski space-time. Fields in the region $x+ t < 0$ are necessarily zero because this domain is not causally connected with the particle's world-line. The $\delta$-function part of the fields for the region $x+t = 0$ is omitted in our work, because it will not be needed to calculate the energy flux. Full expressions for the fields can be found in \cite{Boulware:1979qj} or \cite{Lyle}}:
\begin{equation}
(J^{-1})^{\nu}_{\mu} = \begin{pmatrix} \frac{\ch \tau}{\rho} & - \sh \tau \\   \frac{-\sh \tau}{\rho} & \ch \tau \end{pmatrix} \ ,
\end{equation} 
indeed:
\begin{equation}
A_{\mu} = (J^{-1})^{\nu}_{\mu}A^R_{\nu} = \frac{x^2 + y^2 + z^2  -t^2+ a^{-2}}{x^2-t^2} \frac{(x , -t, 0, 0)}{\sqrt{(x^2 + y^2 + z^2  -t^2+ a^{-2})^2 - 4a^{-2}(x^2 -t^2)}} \ .
\end{equation}
To finish with the first part we will calculate the Poynting vector (i.e. $S =T^{0i}$) and the energy flow for both coordinate systems. 

For the Rinler frame we have:
\begin{equation}
S =T_i^0\propto F_{ik} g^{kl} F_{lm} g^{m0} \propto F_{i}^k F_{k}^0 \ ,
\end{equation}
but the magnetic field is zero,  $F_i^k = 0$, hence $ S = 0$. And it follows that the energy flow is zero as well. This once again proves that there is no radiation in this system.

In the Minkowski space-time it is known that:
\begin{equation}
S = \frac{1}{4 \pi} [E \times H] \ .
\end{equation}
Therefore, before calculating $S$ one should calculate $E$ and $H$. The simplest way to do so is to calculate $E^R_i$ at first, and then transform it back to the Minkowski frame. 

\begin{equation}
E^R_i = -\pd_i  A^R_0  \implies 
\begin{cases}
E^R_1 = \frac{-4 \rho a^{-2}(y^2 + z^2 -\rho^2+ a^{-2}) }{[(\rho^2 + y^2 + z^2 + a^{-2})^2 - 4 a^{-2} \rho^2]^{3/2}} \\
E^R_2 =   \frac{8 y a^{-2}\rho^2 }{[(\rho^2 + y^2 + z^2 + a^{-2})^2 - 4 a^{-2} \rho^2]^{3/2}} \\
E^R_3 =   \frac{8 z a^{-2}\rho^2 }{[(\rho^2 + y^2 + z^2 + a^{-2})^2 - 4 a^{-2} \rho^2]^{3/2}} \ . \\ 
\end{cases}
\end{equation}
The tensor we are going to transform is given by:
\begin{equation}
F^R_{\mu\nu} = \begin{pmatrix} 0 & E^R_1 & E^R_2 &   E^R_3 \\ -E^R_1 & 0 & 0 & 0 \\ -E^R_2 & 0 & 0 & 0 \\ -E^R_3 & 0 & 0 & 0\end{pmatrix} .
\end{equation}
To do the transformation we should multiply it by the inverse Jacoby matrices from both sides:
\begin{align}
F_{\mu \nu} = (J^{-1})^{\eta}_{\mu} F^r_{\eta \phi}(J^{-1})^{\phi}_{\nu}  = \nonumber \\ =\begin{pmatrix} \frac{\ch \tau}{\rho} & - \sh \tau & 0&  0 \\ \frac{-\sh \tau}{\rho} &  \ch \tau & 0 & 0 \\0 & 0 & 1 & 0 \\ 0 & 0 & 0 &1\end{pmatrix} \cdot \begin{pmatrix} 0 & E^R_1 & E^R_2 &   E^R_3 \\ -E^R_1 & 0 & 0 & 0 \\ -E^R_2 & 0 & 0 & 0 \\ -E^R_3 & 0 & 0 & 0\end{pmatrix}  \cdot  \begin{pmatrix} \frac{\ch \tau}{\rho}& \frac{-\sh \tau}{\rho} & 0&  0 \\ - \sh \tau  &  \ch \tau & 0 & 0 \\0 & 0 & 1 & 0 \\ 0 & 0 & 0 &1\end{pmatrix}  = \nonumber \\
= \frac{1}{\rho}\begin{pmatrix} 0 & E^R_1 & E^R_2 \ch \tau &   E^R_3 \ch \tau \\ - E^R_1  & 0 & -E^R_ 2 \sh \tau& -E^R_ 3  \sh \tau\\ - E^R_2 \ch \tau  & E^R_ 2 \sh \tau & 0 & 0 \\- E^R_3 \ch \tau  & E^R_3  \sh \tau & 0 & 0\end{pmatrix} \ .
\end{align}
Thus the electric and the magnetic fields in the lab frame are as follows: 
\begin{equation}
E_1 = \frac{E^r_1}{\rho} \ , E_2 = \frac{E^r_2 \ch \tau}{\rho}  \ ,  E_3 =  \frac{E^r_3 \ch \tau}{\rho}  \ , H_1 = 0 \ , H_2 = \frac{- E^r_3\sh\tau}{\rho} \ , H_3 =  \frac{E^r_ 2 \sh \tau}{\rho} \ ,
\end{equation}
and the Poynting vector components are:
\begin{equation}
S_1= \frac{\ch \tau \sh \tau}{4 \pi \rho^2} [(E^r_2)^2 + (E^r_3)^2]  \ , S_2 =-\frac{ \sh \tau}{4 \pi \rho^2} E^r_1E^r_2 \ , S_3 =-\frac{ \sh \tau}{4 \pi \rho^2} E^r_1E^r_3 \ .
\end{equation}
Using (23) we obtain:
\begin{align}
S_1 = \frac{16 a^{-4} \rho^2(y^2 + z^2 )\ch \tau \sh \tau }{\pi [(\rho^2 + y^2 + z^2 + a^{-2})^2 - 4 a^{-2} \rho^2]^{3}} \ , \nonumber \\
S_2 = \frac{8 a^{-4} \rho y (y^2 + z^2 -\rho^2 + a^{-2})\sh \tau }{\pi [(\rho^2 + y^2 + z^2 + a^{-2})^2 - 4 a^{-2} \rho^2]^{3}} \ ,  \\ 
S_2 = \frac{8 a^{-4} \rho z (y^2 + z^2 -\rho^2 + a^{-2})\sh \tau }{\pi [(\rho^2 + y^2 + z^2 + a^{-2})^2 - 4 a^{-2} \rho^2]^{3}} \ . \nonumber 
\end{align}
And, finally, in terms of $x$ and $t$:
\begin{align}
S_1 = \frac{16 a^{-4} xt(y^2 + z^2 ) }{\pi [(x^2 + y^2 + z^2 + a^{-2}-t^2)^2 - 4 a^{-2} (x^2 -t^2)]^{3}} \ ,\nonumber \\
S_2 = \frac{8 a^{-4}  yt  (y^2 + z^2 +t^2 + a^{-2}-x^2) }{\pi [(x^2 + y^2 + z^2 + a^{-2}-t^2)^2 - 4 a^{-2} (x^2 -t^2)]^{3}} \ , \\ 
S_2 = \frac{8 a^{-4}  zt (y^2 + z^2 +t^2 + a^{-2}-x^2) }{\pi [(x^2 + y^2 + z^2 + a^{-2}-t^2)^2 - 4 a^{-2} (x^2 -t^2)]^{3}} \ .\nonumber 
\end{align}
Now, to prove that the above Poynting vector indeed creates radiation we will calculate the energy flow through the infinitely distant surface $\Omega$, described by\footnotemark[4] $t = R , x^2 + y^2 + z^2 = R^2$ with $R \to \infty$\footnotemark[5]. We can parametrize it with the use of the spherical coordinates $x = R\cos (\theta), y = R\sin(\theta)cos(\phi), z = R \sin(\theta) sin(\phi) $. After the substitution in (29):
\footnotetext[4]{Usually the surface of constant $r = [(x^2 + y^2 + z^2 + a^{-2}-t^2)^2 - 4 a^{-2} (x^2 -t^2)]$ is used, and the resulting energy flux is equal to $\frac{2}{3}a^2$, as in the \cite{Lyle}.}
\footnotetext[5]{We cannot use an infinitely distant surface of finite time, because the radiation created by the particle needs time to reach the surface, hence, while taking the limit we should keep the retarded time $u = t-r$ fixed rather than $t$ itself.}
\begin{align}
S_1 = \frac{16 a^{-4} R^4 \cos(\theta) \sin^2(\theta) }{\pi [a^{-4} + 4 a^{-2}R^2\sin^2(\theta)]^{3}} \ , \nonumber \\
S_2 = \frac{8 a^{-4}R^2 \sin(\theta)\cos(\phi) (2R^2 \sin^2(\theta)+a^{-2}) }{\pi [a^{-4} + 4 a^{-2}R^2\sin^2(\theta)]^{3}} \ ,  \\ 
S_2 = \frac{8 a^{-4}R^2 \sin(\theta)\sin(\phi) (2R^2 \sin^2(\theta)+a^{-2})  }{\pi [a^{-4} + 4 a^{-2}R^2\sin^2(\theta)]^{3}} \nonumber 
\end{align}
To express the flow we need to calculate a scalar product of the Poynting vector and the unit normal vector $\mathbf{n} = (\cos(\theta), \sin(\theta)\cos(\phi), \sin(\theta)\sin(\phi))$:
\begin{equation}
(\mathbf{S}, \mathbf{n}) = \frac{8R^2a^{-4}\sin^2(\theta)(2R^2 + a^{-2}) }{\pi [a^{-4} + 4 a^{-2}R^2\sin^2(\theta)]^{3}} \mathrel{\mathop{\approx}\limits_{R \to \infty}} \frac{16R^4a^{-4}\sin^2(\theta) }{\pi [a^{-4} + 4 a^{-2}R^2\sin^2(\theta)]^{3}} \ .
\end{equation}
Now the energy flow is equal to:
\begin{equation}
J = \int_{\Omega} (\mathbf{S}, d\mathbf{n}) = \int_0^{\pi}d\theta \int_0^{2\pi}d\phi R^2\sin(\theta) (\mathbf{S}, \mathbf{n}) = \int_0^{\pi}d\theta \frac{32R^6a^{-4}\sin^2(\theta) }{ [a^{-4} + 4 a^{-2}R^2\sin^2(\theta)]^{3}}  \ .
\end{equation}
If $R\theta a >> 1$ the function under integration takes the form $\frac{a^2}{2 \sin^3(\theta)}$, and if $R\theta a  << 1$ -- $32R^6 \theta^2 a^8$. So, the whole integral is definitely non-zero, i.e. the radiation is present in the Minkowky space-time.
\subsection{Discussion}
From the above observations  one can conclude that the  energy flow is non-zero in the Minkowski space-time. It can also be seen from the above equalities that if one makes a Lorentz boost so that the particle's velocity is equal to zero (instantly co-moving inertial frame), one gets the expressions for the fields like in the Minkowski frame at $t = 0$ (see (20) and (29)). So, at this single moment of time the radiation is absent from the system and the magnetic  field is zero. Perhaps, that may explain the reason why radiation friction force seems to be zero for a homogeneous acceleration.

Now we can suggest an interpretation of the results obtained in the Rindler frame case. One can see that the magnetic field is vanishing and the electric field is static, so there is no energy flow from the particle for all points with $\rho > 0$. But it should lose energy in both cases, because the energy loss is an invariant under covariant transformations. So we ought to somehow explain this discrepancy. 

One can see that the Rindler coordinate system does not cover the whole space-time, and that there exists a light-like surface $\rho = 0$ which corresponds to the cone $x = t , x \ge 0$ in the Minkowski coordinates. The static coordinate system is not extendable beyond this surface and the radiation is created outside of this coordinate patch.
I.e. one can conclude that the particle does not create radiation near its position, but creates it only outside of the Rindler frame\footnotemark[6]. It is probably worth stressing that the characteristic wavelength of the radiation due to the homogeneously accelerated particle is $\frac{1}{a}$ and, hence, the wave zone starts outside of the Rindler horizon. 
\footnotetext[6]{The similar argument can be found in the reference \cite{newref1}. This paper also contains a standard derivation of the electromagnetic fields for the hyperbolic motion in both frames analogous to the analysis done here.}
We clarify this situation further in the section 4 using waves as sources of radiation in the frames which have been used above. 

\section{Uniformly rotating particle}
It is also interesting to study the situation with another uniformly moving particle as a source of radiation and see if this case is similar to the previous one. Thus, in this section we study a uniformly rotating charged particle.

First, we again describe the coordinate systems, which are  used below. Second, we express the solutions implicitly in both systems\footnotemark[7]. In the end we examine the asymptotic behaviour of the electromagnetic field near the particle's position in the co-moving non-inertial frame.
\footnotetext[7]{We were not able to obtain the explicit form in this case because the resulting equation is transcendental.}
\subsection{Introducing coordinate frames}
In the beginning, as previously, we describe the non-inertial co-moving frame for a uniformly rotating particle, i.e. a particle with the world-line given by:
\begin{equation}
z^0 = \tau, z^1 = R \cos (\omega \tau), z^2 = R \sin (\omega \tau) , z^3 = 0.
\end{equation}
To find the non-inertial co-moving reference system we shall make a transition $x, \ y \to r, \phi$, using:
\begin{equation}
\begin{cases}
x = R \cos(\omega t) + r \cos(\omega t + \phi) \\
y = R \sin(\omega t) + r \sin(\omega t + \phi)
\end{cases}
\Leftrightarrow
\begin{cases}
r = \sqrt{(x - R\cos(\omega t))^2 + (y - R \sin(\omega t))^2} \\
\phi = \arctg \left[ \frac{y - R \sin(\omega t)}{x - R \cos (\omega t)} \right] - \omega t  \ .
\end{cases}
\end{equation}
It is easy to see that in the above coordinates the world-line of our particle is given by: $z^0 = \tau, z^1 = 0, z^ 2 = 0 , z^3 = 0 $, i.e. this system indeed is a co-moving frame.

The coordinate differentials are expressed as follows:
\begin{align}
dx = dr \cos (\phi + \omega t) -  d \phi \cdot r \sin (\phi + \omega t)  - dt  \cdot \omega (R \sin (\omega t) + r \sin (\phi + \omega t))  \nonumber \\ 
dy = dr \sin (\phi + \omega t) +  d \phi\cdot  r \cos (\phi + \omega t)  + dt \cdot   \omega (R \cos (\omega t) + r \cos (\phi + \omega t)) \ .
\end{align}
After substituting them into the formula for the Minkowskian interval we obtain the metric tensor for the co-moving frame:
\begin{multline}
ds^2 = dt^2 - dx^2 - dy^2 -dz^2 = \\
= dt^2 [1 - \omega^2(R^2 + r^2  + 2rR \cos(\phi))] - dr^2 - d \phi^2 r^2 - dz^2 - \\ - 2 dr dt \cdot \omega R \sin (\phi) - 2 d\phi dt \cdot r\omega (r + R \cos (\phi)) \ .
\end{multline}
Or in the matrix form:
\begin{equation}
g_{ij} = \begin{pmatrix} [1 - \omega^2(R^2 + r^2  + 2rR \cos(\phi))] & -\omega R \sin \phi & -\omega r (r + R \cos \phi) & 0 \\ -\omega R \sin \phi & -1 & 0 & 0 \\- \omega r (r + R \cos \phi) & 0 & -r^2 & 0 \\ 0 & 0 & 0 & -1 \end{pmatrix} \ .
\end{equation}
$\dg$ for this metric is equal to $r$, and its inverse can be written as:
\begin{equation}
g^{ij} = \begin{pmatrix} 1 & -\omega R \sin \phi & -\omega  (1 + \frac{R}{r} \cos \phi) & 0 \\ -\omega R \sin \phi & \omega^2R^2\sin^2 \phi -1 & \omega^2 \frac{R}{r} \sin \phi (r+ R \cos \phi)& 0 \\ -\omega  (1 + \frac{R}{r} \cos \phi) & \omega^2 \frac{R}{r} \sin \phi (r+ R \cos \phi) &\frac{1}{r^2} \left [  \omega^2 (r+ R \cos \phi)^2  - 1\right  ]& 0 \\ 0 & 0 & 0 & -1 \end{pmatrix} \ .
\end{equation}
\subsection{Expressing vector and electromagnetic field solutions }
We have introduced our coordinate systems; now it is time to use the Liénard-Wiechert potentials as we did before\footnotemark[8].
\footnotetext[8]{We won't be rewriting the Maxwell equations for this system explicitly as we did in (8) for the Rindler frame, because the resulting equations are too clumsy.}
Let $(t,x,y,z)$ be, as previously, the point of space-time in which the fields are measured, then the velocity three-vector and the three-vector $\mathbf L^i = x^i - z^i$ have the following form ($z^i$ is defined in (33)):
\begin{align}
\mathbf{v} = (-\omega R \sin (\omega \tau), \omega R \cos(\omega t), 0) \nonumber \\
\mathbf{L} = (x - R \cos (\omega \tau), y - R \sin (\omega \tau), z) \ .
\end{align}
Then, as we already know:
\begin{equation}
\begin{cases}
A_{\mu} = \left( \frac{1}{|\mathbf{L}| - (\mathbf{v},\mathbf{L})}   , \frac{- \mathbf{v}}{|\mathbf{L}| - (\mathbf{v},\mathbf{ L})} \right) \\ 
|\mathbf{L}| = t - \tau \ .
\end{cases}
\end{equation}
From the second equation in (40) it follows that:  $|\mathbf{L}| -(\mathbf{v},\mathbf{L}) = t - \tau + \omega R (x \sin (\omega \tau) - y \cos( \wt))$. And using this, the first equation in (40) transforms into:
\begin{equation}
A_{\mu} = \frac{1}{t - \tau + R\omega (x \sin (\omega \tau) - y \cos (\omega \tau) )} \cdot \begin{pmatrix} 1   \\  R \omega \sin (\omega \tau) \\  -R \omega \cos(\omega \tau)\\  0 \end{pmatrix} \ .
\end{equation}
Now let us express (40) in terms of $r$ and  $\phi$. Defining $\psi = \omega (t- \tau)$, we get:
\begin{align}
 t - \tau + R\omega (x \sin (\omega \tau) - y \cos (\omega \tau) ) = t - \tau - R\omega [r \sin \big( \omega (t - \tau) + \phi \big) + R \sin (\omega (t - \tau)) ] = \nonumber \\
= \frac{\psi}{\omega} - Rr\omega \sin(\psi + \phi) - R^2 \omega \sin (\psi) \ ,
\end{align}
and:
\begin{align}
|\mathbf{L}|^2 = (x - R \cos(\wt))^2 + (y - R \sin(\wt))^2 + z^2 = \nonumber \\
= 2R^2 + r^2 +z^2 + 2rR \cos (\phi) - 2 R^2 \cos(\psi) - 2Rr \cos(\phi + \psi) = \frac{\psi^2}{\omega^2} \ .
\end{align}
Thus (40) obtains the following form:
\begin{equation}
\begin{cases}
A_{\mu} = \frac{1}{\frac{\psi}{\omega} - Rr\omega \sin(\psi + \phi) - R^2 \omega \sin (\psi)} \cdot \begin{pmatrix} 1   \\  R \omega \sin (\omega \tau) \\  -R \omega \cos(\omega \tau)\\  0 \end{pmatrix} \\
2R^2 + r^2 +z^2 + 2rR \cos (\phi) - 2 R^2 \cos(\psi) - 2Rr \cos(\phi + \psi) = \frac{\psi^2}{\omega^2} \ .
\end{cases}
\end{equation}
It is time to transform the vector field to the non-inertial co-moving frame, i.e. $A_{\mu} \to A^c_{\mu}$. One can do so using the Jacobi matrix, which is given by: 
\begin{equation}
J^{\nu}_{\mu} = \frac{\partial x^{\nu}}{\partial x'^{\mu}} = \begin{pmatrix} 1 & -r\omega \sin(\phi + \omega t) - R\omega \sin(\omega t)  & r\omega \cos(\phi + \omega t) + R\omega \cos(\omega t)&  0 \\ 0 & \cos (\phi + \omega t)& \sin (\phi + \omega t) & 0 \\0 & -r\sin (\phi + \omega t)& r\cos (\phi + \omega t) & 0 \\ 0 &0 & 0 &1\end{pmatrix}  \ .
\end{equation}
Thus:
\begin{equation}
\begin{cases}
A^c_i = \frac{1}{\frac{\psi}{\omega} - Rr\omega \sin(\psi + \phi) - R^2 \omega \sin (\psi)} \cdot \begin{pmatrix} 1 - Rr \omega^2 \cos (\phi + \psi) - R^2 \omega^2 \cos(\psi)  \\  -R \omega \sin (\phi + \psi) \\  -rR \omega \cos(\phi + \psi)\\  0 \end{pmatrix}  \\
F(\psi, \phi, z, r) = 2R^2 + r^2 +z^2 + 2rR \cos (\phi) - 2 R^2 \cos(\psi) - 2Rr \cos(\phi + \psi) - \frac{\psi^2}{\omega^2} = 0 \ .
\end{cases}
\end{equation}
Here $\psi (r,z,\phi)$ is defined implicitly by the equation $F(\psi, \phi, z, r) = 0$. To calculate the electromagnetic field one needs to take the partial derivatives of $\psi$. In order to do so one should, at first, calculate the partial derivatives of $F$ and then use the implicit function theorem.

Derivatives of $F$ are given by:
\begin{align}
\partial_z F = 2z \ , \partial_r F = 2r + 2R\cos(\phi) - 2 R \cos(\psi + \phi) \ , \partial_{\phi} F = -2 rR \sin(\phi) + 2rR \sin(\phi+ \psi) \ ,\nonumber \\
\partial_{\psi}F = -\frac{2\psi}{\omega^2} + 2R^2 \sin(\psi) + 2rR \sin (\psi + \phi) \ .
\end{align}
And $\psi$ derivatives by:
\begin{align}
\partial_{\phi} \psi = -\frac{\partial_{\phi}F}{\partial_{\psi}F} = -\frac{- rR \sin(\phi) + rR \sin(\phi+ \psi)}{-\frac{\psi}{\omega^2} + R^2 \sin(\psi) + rR \sin (\psi + \phi)} \ , \nonumber \\
\partial_{z} \psi = -\frac{\partial_{z}F}{\partial_{\psi}F} = -\frac{z}{-\frac{\psi}{\omega^2} + R^2 \sin(\psi) + rR \sin (\psi + \phi)} \ , \\
\partial_{r} \psi = -\frac{\partial_{r}F}{\partial_{\psi}F} = -\frac{r + R\cos(\phi) -  R \cos(\psi + \phi) }{-\frac{\psi}{\omega^2} +R^2 \sin(\psi) + rR \sin (\psi + \phi)} \ . \nonumber
\end{align}
Now, one can use  (46) and (48) to find an exact form for the electric and the magnetic fields\footnotemark[9], but we refrain from this here since the resulting formulas are far too complex. Nonetheless, we can make a definite  conclusion about the radiation without the explicit form of the electromagnetic fields. 
\footnotetext[9]{However, $\psi$ is going to be present in those expressions, so they are still going to be implicit.}
\subsection{Asymptomatic behaviour of the electromagnetic field in the co-moving frame}
We have failed to find the exact solution; instead we can find the asymptomatic form of the implicit solution we have found.
We aim to examine its asymptotic behaviour near the particle's position, i.e. $r = 0, z = 0$. In order to do so, we replace $z$ with $\chi \cdot r$, where $\chi \approx 1$, so that all spatial coordinates are of the same order of magnitude. Furthermore, we look  for $\psi$ in the form of $\psi = r \epsilon$. Substituting this into the last formula in (46), and leaving out everything up to the second order in $r$ (first non-vanishing order), we get:
\begin{align}
2R^2 + r^2 + \chi^2 r^2 + 2rR \cos \phi - 2R^2\left(1 - \frac{\epsilon^2 r^2}{2}\right) - 2Rr(\cos \phi - \sin \phi \cdot \epsilon r) - \frac{\epsilon^2r^2}{\omega^2} = 0\implies \nonumber \\
\implies 1 + \chi^2 + \sin \phi \epsilon + \epsilon^2 (R^2 - \frac{1}{\omega^2}) = 0 \ .
\end{align}
Thus $\epsilon$ is defined by a quadratic equation, but from (41) $\psi = |L| \omega \implies \psi = r \epsilon > 0$, so we should take the positive root. Also, we notice that $\omega R$ is equal to the velocity of the particle, so it is lesser than 1. Therefore $\epsilon$ is equal to:
\begin{equation}
\frac{\epsilon}{\omega}  =  \frac{\omega R \sin \phi + \sqrt{(1 + \chi^2)-\omega^2 R^2 (\cos^2 \phi + \chi^2) }}{1 - \omega^2 R^2} \ .
\end{equation}
In this approximation the vector field can be rewritten as follows:
\begin{equation}
A^c_{\mu} = \frac{1}{r[\frac{\epsilon}{\omega} - R \omega \sin \phi - R^2 \omega \epsilon]} \begin{pmatrix} 1 - R^2 \omega^2  \\ -R \omega \sin \phi \\ -r R \omega \cos \phi \\ 0 \end{pmatrix} \ .
\end{equation}
Now we are able to calculate $F_{\mu \nu}$ components with the use of the following formulas:
\begin{equation}
E^c_1 = -\partial_r A^c_0 \ , \ \ E^c_{2i} = - \partial_{\phi} A^c_0  \ , \ \ E^c_3 = - \partial_z A^c_0 \ , \ \ B^c_1 = \partial_z A^c_2 \ , \ \ B^c_2 = - \partial_z A^c_1 \ , \ \ B^c_3 = \partial_{\phi} A^c_1 - \partial_r A^c_2 \ .
\end{equation}
If one defines $Q \approx 1$ as: 
\begin{equation}
Q(\chi, \phi) = \frac{ (1 - R^2 w^2) }{[  R^2 \omega^2 \cos^2\phi -  \chi^2(1- R^2 \omega^2)  -1 ]\sqrt{(  1 + \chi^2 - R^2 \omega^2 \chi^2) -  R^2 \omega^2 \cos^2 \phi}} \ ,
\end{equation}
then one gets the expressions for the electromagnetic fields in the form of:
\begin{align}
E^c_1 =&  \frac{Q(  R^2 \omega^2 \cos^2\phi - 1)}{r^2} \ , \ \ 
E^c_2 = \frac{Q R^2 \omega^2  \cos \phi \sin \phi}{r} \ , \ \ 
E^c_3 = \frac{- Q\chi(1 - R^2 w^2) }{r^2} \ ,
\nonumber \\
B^c_1 =& \frac{- Q \chi R \omega  \cos \phi}{r} \ , \ \ \ \ \ \ \ \ \
B^c_2 = \frac{Q \chi R \omega  \sin \phi}{r^2}  \ , \ \ \ \ \  \ \ \ \ 
B^c_3 =\frac{ Q\chi R \omega  \cos \phi}{r} \ .
\end{align}
The energy-momentum tensor can be calculated explicitly, but the resulting expressions are too complex. However, it can be stated that all of its components are not zero.

\subsection{Discussion}
Once again we see that the radiation is present in the Minkowki space-time\footnotemark[10]. Furthermore, this time  in  the co-moving frame of reference the fields are static, but the energy-momentum tensor components are non-zero nonetheless. However, this coordinate patch also possesses a light-like surface at $\omega (R+r) = 1$, beyond which the static metric under consideration cannot be extended. Thus, the radiation cannot be defined as an energy flow through an infinitely distant surface because there is none. Moreover, the characteristic wavelength of the radiation is of the order of the distance between the source and the light-like surface mentioned above, hence, the wave-zone starts beyond the region of space-time covered by the co-moving frame.

The similar analysis for the general situation of an arbitrary accelerated particle can be found in the reference \cite{newref2}. In this article the approximate form of the fields in a neighborhood of the source are found. However, alike the results for the asymptotic form for the fields of uniformly rotating particle obtained in the section above, this analysis is not useful for the study of radiation, due to its locality. 
\footnotetext[10]{This is a well known result.}
\section{Wave as a source of radiation}
We have discussed the radiation in different coordinate systems for point-like sources. But it is also interesting to study other types of sources. Now we proceed to discuss the situation where a scalar field excitation is used to create radiation, i.e. the vector field harmonics.
First, we  introduce the action and formulate the  problem under consideration  more rigorously. Second, we consider the problem for a source moving with zero acceleration in the Minkowski space-time to clarify our methods in a simple setting. Third, we solve the same problem for the Rindler frame. Fourth, we investigate the case of uniformly accelerated harmonics in the Minkowski space-time. In the end we discuss the results and examine the connection between this case and the case of the point-like source.
A similar situation was considered in the de Sitter space in  \cite{Akhmedov:2009be} and \cite{Akhmedov:2008pu}. 
\subsection{Introducing the action}
The theory which we discuss here describes the interaction between a scalar field and a gauge vector field:
\begin{equation}
S = \int d^4x \sqrt{-g} \left[ -\frac{1}{4} F^{\mu \nu} F_{\mu \nu} + (D[A]_\mu \phi)^{*} (D[A]^\mu \phi) - m^2 \phi^{*} \phi \right] \ ,
\end{equation}
here  $D[A]_\mu = \partial_\mu - ie A_\mu$. The coupling constant $e$ is zero at the  past infinity, then it increases adiabaticly at the distant past, stays constant for a long while and, afterwards, is adiabaticly switched off at the distant future to be zero again at the future infinity. We assume that the time derivative of $e$ is much smaller than $e$ itself, so it is omitted. 

Below the equations of motions for both scalar and gauge fields are written and the Lorentz gauge for the vector field is assumed:
\begin{equation}
\begin{cases}
\Big [m^2 + \frac{1}{\sqrt{-g}}D[A]_\mu (\sqrt{-g} D[A]^\mu) \Big]\phi = 0 \\
g_{\mu \eta} \frac{1}{\sqrt{-g}} \partial_\nu \left( \sqrt{-g} F^{\nu \mu} \right) = -ie(\phi^* \partial_\eta \phi - \phi \partial_\eta \phi^*)- 2e^2A_\eta \phi \phi^*\\
g^{\mu \nu} \partial_\mu A_\nu = g^{\mu \nu} \Gamma^\eta_{\mu \nu} A_\eta \ .
\end{cases}
\end{equation}
We discuss the following process: there is a scalar field excitation and no vector field on the past infinity, and then, after we switch the interaction on, the scalar field can possibly create a vector field excitation. We are interested in the resulting vector field in the points of space-time with $x^0$ after the coupling constant is turned off. When the interaction is switched off the vector field should satisfy the homogeneous equations, and, thus, should be expressed in terms of the harmonics. In the calculations below, we assume that the coupling constant is much smaller than the amplitude of the scalar field ($e << \phi$) , therefore, we are interested in the solution at the linear order in $e$. Hence, we approximate:
\begin{equation}
\begin{cases}
D[A]_\mu \approx \partial_\mu \\
-ie(\phi^* \partial_\eta \phi - \phi \partial_\eta \phi^*)- 2e^2A_\eta \phi \phi^* \approx -ie(\phi^* \partial_\eta \phi - \phi \partial_\eta \phi^*) \ ,
\end{cases} 
\end{equation}
and the system of equations (56) is transformed into:
\begin{equation}
\begin{cases}
\Big [m^2 + \frac{1}{\sqrt{-g}}\pd_\mu (\sqrt{-g} \pd^\mu) \Big]\phi = 0 \\
g_{\mu \eta} \frac{1}{\sqrt{-g}} \partial_\nu \left( \sqrt{-g} F^{\nu \mu} \right) \approx -ie(\phi^* \partial_\eta \phi - \phi \partial_\eta \phi^*)  \\
g^{\mu \nu} \partial_\mu A_\nu \approx g^{\mu \nu} \Gamma^\eta_{\mu \nu} A_\eta \ .
\end{cases}
\end{equation}
\subsection{Free moving source in the Minkowski space-time}
We start by solving our problem in the Minkowski space-time. For this frame the equations under consideration are much simpler:
\begin{equation}
\begin{cases}
 [m^2 + \pd_\mu \pd^\mu ]\phi = 0 \\
  \partial_\nu   F^{\nu }_{\mu}  = -ie(\phi^* \partial_\mu \phi - \phi \partial_\mu \phi^*)\\
 \partial^\mu A_\mu = 0 \ .
\end{cases}
\end{equation}
The scalar field harmonics in this case are given by  $e^{i p x}$, where $p =p_\mu$ is a four-vector $p_\mu = (\sqrt{m^2 + p^2}, \mathbf{p})$, $\mathbf{p}$ is a three-vector and $px$ stands for a four-convolution $p_\mu x^\mu$. Hence, an arbitrary scalar field can be expanded in the following form:
\begin{equation}
\phi(x) = \int_{\mathbb{R}^3} d^3  \mathbf p   \ a_{\mathbf p} e^{i p x} \ .
\end{equation}
With this choice of $\phi$ we can write down the expression for the current as follows:
\begin{gather}
j = -ie \left[ \int d^3 \mathbf p_1 d^3 \mathbf p_2 \ a^*_{\mathbf p_1} e^{-i p_1 x} \ a_{\mathbf p_2} i p_2  e^{i p_2 x} +  \int d^3 \mathbf p_1 d^3 \mathbf p_2 \ a_{\mathbf p_1} e^{i p_1 x} \ a^*_{\mathbf p_2} i p_2  e^{-i p_2 x}  \right] = \nonumber \\ = e \int d^3 \mathbf p_1 d^3 \mathbf p_2 \ (p_1 + p_2) a^*_{\mathbf p_1}a_{\mathbf p_2}e^{i (p_2 -p_1) x} \ .
\end{gather}
Using the Lorentz gauge we transform the equations for the vector field to be of the following form:
\begin{equation}
\Box A_\mu (x)= j_\mu(x) \ .
\end{equation}
In order to solve these equations we can use the Green's function, defined by $\Box G(x,y) = \delta^{(4)}(x-y)$; indeed, it is easy to check that:
\begin{equation}
A_\mu (x)= \int_{\mathbb{R}^4} d^4 y G(x,y) j_\mu(y) \ ,
\end{equation}
is a solution. We want to represent $A_\mu$ as a sum of the vector field harmonics\footnotemark[11]. In order to do so we shall Fourier expand the Green's function:
\footnotetext[11]{For the flat space-time they are given by $e^{ik_\mu x^\mu}$ with $k_\mu k^\mu = 0$.}
\begin{multline}
\Box \int_{\mathbb {R}^4} \frac{d^4 k}{(2 \pi)^4} e^{i k_\mu x^\mu} \tilde{G}(k,y) = \int_{\mathbb {R}^4} \frac{d^4 k}{(2 \pi)^4} e^{i k_\mu x^\mu - i k_\mu y^\mu} \implies \\ \implies \int_{\mathbb {R}^4} \frac{d^4 k}{(2 \pi)^4} \big[\Box e^{i k_\mu x^\mu} \big]\tilde{G}(k,y) = \int_{\mathbb {R}^4} \frac{d^4 k}{(2 \pi)^4} e^{i k_\mu x^\mu - i k_\mu y^\mu} \implies  \\ \implies \int_{\mathbb {R}^4} \frac{d^4 k}{(2 \pi)^4} \big[- k_\mu k^\mu e^{i k_\mu x^\mu} \big]\tilde{G}(k,y) = \int_{\mathbb {R}^4} \frac{d^4 k}{(2 \pi)^4} e^{i k_\mu x^\mu - i k_\mu y^\mu} \implies \tilde{G}(k,y) = \frac{e^{-i k_\mu y^\mu}}{- k_\mu k^\mu} \ .
\end{multline}
Here we used the fact that $\Box$ acts only on $x$ and the Fourier transformation is reversible. Thus:
\begin{multline}
G(x,y) = \int_{\mathbb {R}^4} \frac{d^4 k}{(2 \pi)^4} e^{i k_\mu x^\mu} \tilde{G}(k,y) = \int_{\mathbb {R}^4} \frac{d^4 k}{(2 \pi)^4} \frac{e^{i k_\mu (x^\mu-y^\mu)}}{- k_\mu k^\mu}  = \int_{\mathbb{R}^3}\frac{d^3 \mathbf k}{(2 \pi)^3}  e^{ i (\mathbf{k}, \mathbf{x}-\mathbf{y})}\int_{\mathbb{R}}\frac{dk_0}{2 \pi}\frac{e^{i k_0 (x^0-y^0) }}{|\mathbf{k}|^2 - k_0^2} \ .
\end{multline}
Here $\mathbf{k} = (k_1,k_2,k_3)$ and $\mathbf{x} = (x^1,x^2,x^3)$.

In our problem the Green's function also needs to be of the retarded form:
\begin{multline}
G(x,y) = \int_{\mathbb{R}^3}\frac{d^3 \mathbf k}{(2 \pi)^3}  e^{ i (\mathbf{k}, \mathbf{x}-\mathbf{y})}\int_{\mathbb{R}}\frac{dk_0}{2 \pi}\frac{e^{i k_0 (x^0-y^0) }}{(|\mathbf k| - k_0)(|\mathbf k| + k_0)} =\\= -\theta (x^0 -y^0) \int_{\mathbb{R}^3}\frac{d^3 \mathbf k}{(2 \pi)^3}  e^{ i (\mathbf{k}, \mathbf{x}-\mathbf{y})} i\left [ \frac{e^{i |\mathbf k| (x^0-y^0)}}{2|\mathbf k|} -\ \frac{e^{-i |\mathbf k| (x^0-y^0)}}{2|\mathbf k|} \right] = \\ = -\theta (x^0 -y^0)\int_{\mathbb{R}^3}\frac{d^3 \mathbf k}{(2 \pi)^3} \frac{i}{2|\mathbf k|} \left [ e^{i |\mathbf k| (x^0-y^0) + i (\mathbf{k}, \mathbf{x}-\mathbf{y}) } - e^{-i |\mathbf k| (x^0-y^0) + i (\mathbf{k}, \mathbf{x}-\mathbf{y}) }\right] \ .
\end{multline}
We calculate the vector field by the substitution of (66) and (61) into (63): 
\begin{gather}
A(x) = -\int \frac{d^3 \mathbf k}{(2 \pi)^3} d^3 \mathbf p_1 d^3 \mathbf p_2 d^3 \mathbf y dy^0 \frac{ie(y^0)}{2|\mathbf k|} e^{ i (\mathbf{k}, \mathbf{x}-\mathbf{y})}\theta (x^0 -y^0)   (p_1 + p_2) a^*_{\mathbf p_1}a_{\mathbf p_2}e^{i (p_2 -p_1) y}\left [ e^{i |\mathbf k| (x^0-y^0)} -c.c. \right] \ .
\end{gather}
After integrating over $\mathbf y$ we get the momentum conservation law in the form of $\delta$-function:
\begin{gather}
\int \frac{dy^3}{(2 \pi)^3} e^{i(- \mathbf k + \mathbf p_2 - \mathbf p_1 ,\mathbf y)} = \delta^{(3)} ( \mathbf p_2 - \mathbf p_1- \mathbf k ) .
\end{gather}
Now we would like to fetch the energy conservation law from the integration over $y^0$. This integral is of the form:
\begin{gather}
\int dy^0 e(y^0) \theta (x^0-y^0) e^{i(p_2-p_1)_0 y^0}\left[ e^{i |\mathbf k| (x^0-y^0)} -c.c. \right] \ .
\end{gather}
To calculate this we can argue as follows.
First, we consider the  point $x$ to be after the time when $e$ is adiabatically switched off, so we can discard the $\theta$-function. Second, the $e$ switches on and off adiabatically at  plus and minus infinities, so in the limit we can take it out from the integral to obtain:
\begin{gather}
\lim_{T \to \infty}e \int_{-T}^{T} dy^0 e^{i(p_2-p_1)_0 y^0}\left[ e^{i |\mathbf k| (x^0-y^0)} -c.c. \right] = 2 \pi e [e^{i |\mathbf k| x^0}\delta ((p_2 -p_1)_0 - |\mathbf k|) - e^{-i |\mathbf k| x^0}\delta ((p_2 -p_1)_0 + |\mathbf k|) ] \ .
\end{gather}
This argument can be made rigorous with the exact calculation, for example, one can take the coupling constant to be equal to $e(y^0) = e \cdot e^{-\varepsilon (y^0)^2}$ and in the limit $\varepsilon \to 0$ and $x^0 \to \infty$ obtain (70).

Thus, we obtained the following expression under the integral in (67):
\begin{gather}
 \delta^{(3)} ( \mathbf p_2 -  p_1- \mathbf k ) [e^{i |\mathbf k| x^0}\delta ((p_2 -p_1)_0 - |\mathbf k|) - e^{-i |\mathbf k| x^0}\delta ((p_2 -p_1)_0 + |\mathbf k|) ] \ .
\end{gather}
But it is easy to see that the conditions imposed by the $\delta$-functions cannot be satisfied simultaneously. Indeed, to fulfill them we should be able to find two points on the mass shell described by the upper half of the hyperboloid $(p_0)^2 = m^2 + \mathbf p^2$ which are connected by a light-like segment\footnotemark[12]. In other words, the radiation by a free floating sources in the   Minkowski space-time is forbidden by the energy-momentum conservation. 
\footnotetext[12]{This segment represents the energy-momentum 4-vector of the photon which is null.}
\subsection{Free moving source in the co-moving Rindler frame}
Now, we repeat the above calculations for the Rindler frame. Equations (58) are transformed as follows:
\begin{equation}
\begin{cases}
\big [m^2 + \Box_R \big]\phi = 0 \\
\big [\Box_{R} + \frac{2\partial_1}{\rho}\big ]A_0 + \frac{2\partial_0}{\rho}A_1 = j_0 \\
\big [\Box_R + \frac{1}{\rho^2}\big]A_1 = j_1 \\
\Box_R A_2 = j_2 \\
\Box_R A_3 = j_3\\
g^{\mu \nu} \partial_\mu A_\nu  = \frac{A_1}{\rho} \ ,
\end{cases}
\end{equation}
here $ \Box_R = \frac{\pd_0\pd_0}{\rho^2} - \frac{\pd_1}{\rho} - \pd_1\pd_1 - \pd_2\pd_2 - \pd_3\pd_3$ and $j_\mu =  -ie(\phi^* \partial_\mu \phi - \phi \partial_\mu \phi^*)$ as before.

For the next step one should find the scalar field harmonics, i.e. the harmonics for the operator $m^2 + \Box_R$. To do so one should use the MacDonald function $K_{\alpha} (px)$, which satisfies the equation: 
\begin{equation}
\left [\partial_x^2 + \frac{\partial_x}{x} - p^2 + \frac{\alpha^2}{x^2}\right ]K_{i\alpha}(px) = 0 \ .
\end{equation}
If we denote:
\begin{equation}
f(\tau,\rho,y,z | \alpha, \omega, p, p_y, p_z) = K_{i\alpha}(p\rho) \cdot e^{ i \omega \tau - i p_y y - i p_z z } \ ,
\end{equation}
operator $m^2 + \Box_R$ acts on it as follows:
\begin{equation}
(m^2 + \Box_{R})f =\left( -p^2 + p_y^2 + p_z^2 + m^2 + \frac{\alpha^2 - \omega^2}{\rho^2}\right)f \ .
\end{equation}
From the above one can conclude that $\alpha = \omega$\footnotemark[13] and $p = \sqrt{m^2 + p_y^2+ p_z^2}$. If we  define $\tilde p \equiv \sqrt{m^2 + \mathbf p^2}$ and use bold letters to represent transversal two-vectors (for example $\mathbf{p}= (p_y, p_z)$), the scalar field can be expanded as\footnotemark[14]:
\footnotetext[13]{The option $\alpha = -\omega$ is irrelevant because $K_{i\alpha}(x) = K_{-i\alpha}(x)$.}
\footnotetext[14]{We do not use the $I_{i\alpha}(x)$ solutions of the modified Bessel equation in the following expansion because their absolute value is not bounded from above at the plus infinity.}
\begin{equation}
\phi = \int_{\mathbb{R_+}} d \omega \int d^2 \mathbf {p} a_{\omega, \mathbf {p}} K_{i \omega} ( \tilde p  \rho) e^{i \omega \tau + i (\mathbf{p}, \mathbf {y})} \ .
\end{equation}
If one uses this to calculate the currents, one gets the following expressions:\footnotemark[15]
\footnotetext[15]{Here we are using the fact that $K_{i\alpha}(x)$ is real. It easily follows from the integral representation of the  MacDonald function: $K_\alpha(x) = \int_0^\infty \exp(-x\cosh t) \cosh(\alpha t)dt$.}
\begin{gather}
j_0 = e \int_{\mathbb{R_+}^2} d\omega_1 d\omega_2 \int d^2 \mathbf{p}_1 d^2 \mathbf p_2  [\omega_1 + \omega_2] a^*_{\omega_1, \mathbf {p}_1}a_{\omega_2, \mathbf {p}_2}  K_{i \omega_1} ( \tilde p_1  \rho)  K_{i \omega_2} ( \tilde p_2  \rho) e^{i (\omega_1-\omega_2) \tau + i (\mathbf{p}_2-\mathbf{p}_1, \mathbf {y})} \ , \nonumber\\
j_1 = 0 \ ,\\ 
\mathbf j = \int_{\mathbb{R_+}^2} d\omega_1 d\omega_2 \int d^2 \mathbf{p}_1 d^2 \mathbf p_2  [\mathbf{p}_1 + \mathbf{p}_2] a^*_{\omega_1, \mathbf {p}_1}a_{\omega_2, \mathbf {p}_2} K_{i \omega_1} ( \tilde p_1  \rho)  K_{i \omega_2} ( \tilde p_2  \rho) e^{i (\omega_1-\omega_2) \tau + i (\mathbf{p}_2-\mathbf{p}_1, \mathbf {y})} \ . \nonumber
\end{gather}
From $j_1=0$ it follows that $A_1$ is the solution of the homogeneous equation, but we are interested only in an inhomogeneous part of the solution, so we can set $A_1 = 0$. As the result we are left with the following equations:
\begin{equation}
\begin{cases}
\left [\Box_{R} + \frac{2\partial_1}{\rho}\right ]A_0 =  j_0  \\
\Box_R \mathbf{A} = \mathbf{j} \ .
\end{cases}
\end{equation}
To solve them we should express the retarded Green's functions for the two operators given in (78) in terms of their harmonics as well. Let us start with the one for the zero component, i.e. $G_0(x|x')$. To find the harmonics one should notice that:
\begin{equation}
\left [\partial_x^2 -\frac{\partial_x}{x} - k^2 + \frac{\alpha^2+1}{x^2} \right ]\big[ k x K_{i\alpha}(kx)\big] = 0 \ ,
\end{equation}
and:
\begin{equation}
\left [\Box_{R} + \frac{2\partial_1}{\rho}\right ] [k \rho \cdot f] = \left [ -k^2 + k_y^2 + k_z^2  + \frac{1 + \alpha^2 -\omega^2}{\rho^2} \right] f \ ,
\end{equation}
with $f$ is defined in (74).

So if we denote $\psi (x | \alpha, \omega, \mathbf k) = |\mathbf k| \rho  \cdot f(\tau,\rho,y,z | \alpha, \omega, |\mathbf{k}|, k_y, k_z) $, with $\mathbf y = (y,z) $ , $\mathbf k = (k_y,k_z)$ and $x = (\tau, \rho, \mathbf y)$, the vector field harmonics are expressed as follows:
\begin{equation}
\psi (x | \alpha, \pm \sqrt{1 + \alpha^2}, \mathbf k) =   |\mathbf k| \rho K_{i\alpha}(|\mathbf{k}|\rho) \cdot e^{\pm  i \sqrt{1 + \alpha^2} \tau - i (\mathbf k, \mathbf y) } \ .
\end{equation}
To express the Green's function in terms of $\psi$ we should use the following identity:
\begin{equation}
\delta (x-x') = \frac{1}{\sqrt{xx'}}\int_{\mathbb R_+} d \alpha \frac{2 \alpha \sh \pi \alpha}{\pi^2} K_{i\alpha}(x)K_{i\alpha}(x') \ .
\end{equation}
With the help of (82) we can obtain an expression for the four dimensional $\delta$-function:
\begin{multline}
\delta^{(4)} (x-x') = \int_{\mathbb R_+} \frac{d \alpha}{\pi^2} \int_{\mathbb R^3}\frac{d\omega d^2 \mathbf{•} k}{(2 \pi)^3} \frac{2 \alpha \sh \pi \alpha}{\mk \rho } K_{i\alpha}(|\mathbf k|\rho)K_{i\alpha}(\mk \rho') e^{i \omega (\tau -\tau') - i (\mathbf k, \mathbf y - \mathbf y')} =\\ = \int_{\mathbb R_+} \frac{d \alpha}{\pi^2} \int_{\mathbb R^3}\frac{d\omega d^2 \mathbf k}{(2 \pi)^3} \frac{2 \alpha \sh \pi \alpha}{\mk^3 \rho^2 \rho' }\psi (x | \alpha, \omega, \mathbf k) \psi (x' | \alpha, -\omega, -\mathbf k) \ .
\end{multline}
We substitute this into the Green's function definition and repeat the steps which we have done for the Minkowski space-time $\left( \text{here} \ \ \mathcal D =\left [\Box_{R} + \frac{2\partial_1}{\rho}\right ] \right)$:
\begin{gather}
 \mathcal D G_0(x|x') = \delta^{(4)} (x-x') \nonumber \\  \Updownarrow \nonumber \\ 
\mathcal D\int_{\mathbb R_+} \frac{d \alpha}{\pi^2} \int_{\mathbb R^3}\frac{d\omega d^2 \mathbf k}{(2 \pi)^3}  \psi (x | \alpha, \omega, \mathbf k) \tilde G_0 (x'| \alpha, \omega, \mathbf k) = \int_{\mathbb R_+} \frac{d \alpha}{\pi^2} \int_{\mathbb R^3}\frac{d\omega d^2 \mathbf k}{(2 \pi)^3} \frac{2 \alpha \sh \pi \alpha}{\mk^3 \rho^2 \rho' } \psi (x | \alpha, \omega, \mathbf k) \psi (x' | \alpha, -\omega, -\mathbf k) \nonumber \\  \Updownarrow \nonumber \\ 
\int_{\mathbb R_+} \frac{d \alpha}{\pi^2} \int_{\mathbb R^3}\frac{d\omega d^2 \mathbf k}{(2 \pi)^3} \left [\mathcal D \psi (x | \alpha, \omega, \mathbf k)\right] \tilde G_0 (x'| \alpha, \omega, \mathbf k) = \int_{\mathbb R_+} \frac{d \alpha}{\pi^2} \int_{\mathbb R^3}\frac{d\omega d^2 \mathbf k}{(2 \pi)^3} \frac{2 \alpha \sh \pi \alpha}{\mk^3 \rho^2 \rho' } \psi (x | \alpha, \omega, \mathbf k) \psi (x' | \alpha, -\omega, -\mathbf k) \nonumber \\  \Updownarrow \nonumber \\ 
\frac{1 + \alpha^2 -\omega^2}{\rho^2}  \tilde G_0 (x'| \alpha, \omega, \mathbf k) = \frac{2 \alpha \sh \pi \alpha}{\mk^3 \rho^2 \rho' }  \psi (x' | \alpha, -\omega, -\mathbf k)\nonumber \\  \Updownarrow \nonumber \\ 
\tilde G_0 (x'| \alpha, \omega, \mathbf k) = \frac{2 \alpha  \sh \pi \alpha}{\mk^3  \rho' } \cdot  \frac{\psi (x' | \alpha, -\omega, -\mathbf k)}{1 + \alpha^2 -\omega^2} \nonumber \\  \Updownarrow \nonumber \\  
G_0(x | x') = \int_{\mathbb R_+} \frac{d \alpha}{\pi^2} \int_{\mathbb R^3}\frac{d\omega d^2 \mathbf k}{(2 \pi)^3} \frac{2 \alpha \sh \pi \alpha}{\mk^3  \rho' }\cdot  \frac{\psi (x | \alpha, \omega, \mathbf k) \psi (x' | \alpha, -\omega, -\mathbf k)}{1 + \alpha^2 -\omega^2} \ .
\end{gather}
As previously, we take the $\omega$ integral using residues:
\begin{gather}
G_0(x | x') =\int_{\mathbb R_+} \frac{d \alpha}{\pi^2} \int_{\mathbb R^2}\frac{d^2 \mathbf k}{(2 \pi)^2} \frac{2 \alpha  \sh \pi \alpha}{\mk^3 \rho'} \int_{\mathbb R} \frac{d \omega}{2 \pi} \frac{\psi (x | \alpha, \omega, \mathbf k) \psi (x' | \alpha, -\omega, -\mathbf k)}{(\beta -\omega)(\beta +\omega)} = \nonumber \\ 
= - \int_{\mathbb R_+} \frac{d \alpha}{\pi^2} \int_{\mathbb R^2}\frac{d^2 \mathbf k}{(2 \pi)^2} \frac{2 \alpha \sh \pi \alpha}{\mk^3  \rho' } \cdot i \left[  \frac{\psi (x | \alpha, \beta, \mathbf k) \psi (x' | \alpha, -\beta, -\mathbf k)- \psi (x | \alpha, -\beta, \mathbf k) \psi (x' | \alpha, \beta, -\mathbf k)}{2\beta}  \right ]\theta(\tau -\tau') \ ,
\end{gather}
with $\beta \equiv \sqrt{1 + \alpha^2}$. Thus, it is expressed in terms of the vector field harmonics. We also would like to rewrite Green's function in the following form, which would be used in the vector field calculations below:
\begin{equation}
G_0(x|x')=- \int_{\mathbb R_+} \frac{d \alpha}{\pi^2} \int_{\mathbb R^2}\frac{d^2 \mathbf k}{(2 \pi)^2} \frac{ i \alpha   \rho  \sh \pi \alpha}{\mk \sqrt{1 + \alpha^2}} K_{i \alpha} (\mk \rho) K_{i \alpha} (\mk \rho') e^{i (\mathbf k, \mathbf y -\mathbf y')}\left[ e^{i\sqrt{1+\alpha^2} (\tau - \tau')} -  c.c \right]\theta(\tau -\tau') \ .
\end{equation}
Now, we shall find the Green's function for the transversal part of the vector field. As we know already, these components satisfy the equation $\Box_R \mathbf A = \mathbf  j$. Let us define:
\begin{equation}
\gamma (x | \omega, \alpha, \mathbf k) = K_{i \alpha} (|\mathbf k| \rho) e^{i \omega \tau + i (\mathbf k, \mathbf y)} \ .
\end{equation}
It is easy to check that these are in fact the harmonics for $\Box_R$.\footnote[16]{To see this one may want to use (75) with $m = 0$.}

If we represent the Green's function as the following integral:
\begin{equation}
G_\perp(x|x') = \int_{\mathbb R_+} \frac{d^+ \alpha}{\pi^2}\int \frac{d\omega d^2\mathbf{k}}{(2\pi)^3} \gamma(x | \alpha, \omega, \mathbf k)\tilde G_\perp (x' | \alpha,\omega, \mathbf k) \ , 
\end{equation}
use (82) and the definition of $G_\perp(x|x')$, we obtain:
\begin{equation}
\tilde G_\perp (x' | \alpha,\omega, \mathbf k) = \frac{2  \alpha  \rho'\sh \pi \alpha}{|\mathbf k| }\frac{\gamma(x' | \alpha, -\omega, -\mathbf k)}{\alpha^2 - \omega^2} \ .
\end{equation}
After we substitute (89) into (88) we get:
\begin{equation}
G_\perp(x|x') = \int_{\mathbb R_+} \frac{d \alpha}{\pi^2}\int \frac{d\omega d^2\mathbf{k}}{(2\pi)^3}\frac{2  \alpha  \rho' \sh \pi \alpha}{|\mathbf k| }\frac{ \gamma(x | \alpha, \omega, \mathbf k)\gamma(x' | \alpha, -\omega, -\mathbf k)}{\alpha^2 - \omega^2}\theta(\tau -\tau') \ .
\end{equation}
Integrating over the poles of $\omega$ to get the retarded form:
\begin{gather}
G_\perp(x|x')  = - \int_{\mathbb R_+} \frac{d \alpha}{\pi^2} \int \frac{ d^2\mathbf{k}}{(2\pi)^2}\frac{i  \rho'\sh \pi \alpha}{|\mathbf k| }\left[ \gamma(x | \alpha, \alpha, \mathbf k)\gamma(x' | \alpha, -\alpha, -\mathbf k) -  \gamma(x | \alpha, -\alpha, \mathbf k)\gamma(x' | \alpha, \alpha, -\mathbf k) \right]\theta(\tau -\tau') = \nonumber \\
= - \int_{\mathbb R_+} \frac{d \alpha}{\pi^2}\int \frac{ d^2\mathbf{k}}{(2\pi)^2}\frac{i  \rho'\sh \pi \alpha}{|\mathbf k| }K_{i\alpha} (|\mathbf k| \rho)K_{i\alpha} (|\mathbf k| \rho') e^{i (\mathbf k, \mathbf y - \mathbf y')}\left[ e^{i\alpha (\tau - \tau')} -  c.c \right]\theta(\tau -\tau') \ .
\end{gather}
Thus, the retarded Green's functions for both operators are obtained and expressed in terms of the corresponding harmonics.
Everything is ready for the calculation of the vector field.
Let us start with its transversal components:
\begin{gather}
\mathbf A = \int_{\mathbb R_+}  d \rho' \int d \tau' d^2 \mathbf y' G_\perp (x|x') \mathbf j(x') = \nonumber \\
=  - \int_{\mathbb R_+^4}  d\omega_1 d\omega_2 d\rho'\frac{d \alpha}{\pi^2} \int d^2 \mathbf{p}_1 d^2 \mathbf p_2 \frac{ d^2\mathbf{k}}{(2\pi)^2} d \tau'  d^2 \mathbf y' \ e [\mathbf{p}_1 + \mathbf{p}_2] a^*_{\omega_1, \mathbf {p}_1}a_{\omega_2, \mathbf {p}_2} \frac{i  \rho'\sh \pi \alpha}{|\mathbf k| } \theta(\tau -\tau') \times \nonumber \\
\times K_{i \omega_1} ( \tilde p_1  \rho')  K_{i \omega_2} ( \tilde p_2  \rho') K_{i\alpha} (|\mathbf k| \rho)K_{i\alpha} (|\mathbf k| \rho') e^{i(\mathbf k, \mathbf y)} e^{i(\mathbf p_2 - \mathbf p_1 - \mathbf k, \mathbf y')}\left[ e^{i\alpha\tau }e^{i(\omega_2 - \omega_1 - \alpha)\tau'} - e^{-i\alpha\tau }e^{i(\omega_2 - \omega_1 + \alpha)\tau'} \right] \ .
\end{gather}
The integral over $ d^2 \mathbf y'$, leads to:
\begin{equation}
\int \frac{d^2 \mathbf y'}{(2 \pi)^2}  e^{i(\mathbf p_2 - \mathbf p_1 - \mathbf k, \mathbf y')} = \delta^{(2)} (\mathbf p_2 - \mathbf p_1 - \mathbf k) \ .
\end{equation}
The  integral over $d \tau'$ with the limit taken as in the previous subsection, gives:
\begin{gather}
\lim_{T \to \infty} \int_{-T}^T \frac{d \tau'}{2 \pi} e(\tau') \theta (\tau - \tau') \left[ e^{i\alpha\tau }e^{i(\omega_2 - \omega_1 - \alpha)\tau'} - e^{-i\alpha\tau }e^{i(\omega_2 - \omega_1 + \alpha)\tau'} \right]= \nonumber \\ = e^{i \alpha \tau} \delta (\omega_2 - \omega_1 -\alpha) - e^{-i \alpha \tau} \delta (\omega_2 - \omega_1 + \alpha) \ .
\end{gather}
But because we integrate  $\alpha$ over $\mathbb R_+$, $\omega_2 -\omega_1$ should be positive in the case $\alpha = \omega_2 - \omega_1$ and negative in the case $\alpha = \omega_1 - \omega_2$. Thus, we can insert $\theta$-functions in the integral beside the corresponding $\delta$-functions. Using all of the above expressions in (92), we get:
\begin{gather}
\mathbf A
=  - \int_{\mathbb R_+^4}  d\omega_1 d\omega_2 d\rho' \frac{d \alpha}{\pi} \int d^2 \mathbf{p}_1 d^2 \mathbf p_2  d^2\mathbf{k} \  [\mathbf{p}_1 + \mathbf{p}_2] a^*_{\omega_1, \mathbf {p}_1}a_{\omega_2, \mathbf {p}_2} \frac{2 i  \rho'\sh \pi \alpha}{|\mathbf k| }  K_{i \omega_1} ( \tilde p_1  \rho')  K_{i \omega_2} ( \tilde p_2  \rho') K_{i\alpha} (|\mathbf k| \rho)K_{i\alpha} (|\mathbf k| \rho') \times \nonumber \\
\times e^{i(\mathbf k, \mathbf y)}\delta^{(2)} (\mathbf p_2 - \mathbf p_1 - \mathbf k)\left[ e^{i \alpha \tau} \delta (\omega_2 - \omega_1 -\alpha) \theta(\omega_2 - \omega_1) - e^{-i \alpha \tau} \delta (\omega_2 - \omega_1 + \alpha)  \theta(\omega_1 - \omega_2)\right] \ .
\end{gather}
Now, if one  renames $\omega_2$ as $\omega_1$ and vice versa in the second summand of the integral, one obtains:
\begin{gather}
\mathbf A
=  - \int_{\mathbb R_+^4}  d\omega_1 d\omega_2 d\rho'\frac{d \alpha}{\pi} \int d^2 \mathbf{p}_1 d^2 \mathbf p_2  d^2\mathbf{k}  \  [\mathbf{p}_1 + \mathbf{p}_2] [a^*_{\omega_1, \mathbf {p}_1}a_{\omega_2, \mathbf {p}_2} - a_{\omega_1, \mathbf {p}_1}a^*_{\omega_2, \mathbf {p}_2}] \frac{2 i  \rho'\sh \pi \alpha}{|\mathbf k| } \times \nonumber \\
\times  K_{i \omega_1} ( \tilde p_1  \rho')  K_{i \omega_2} ( \tilde p_2  \rho') K_{i\alpha} (|\mathbf k| \rho)K_{i\alpha} (|\mathbf k| \rho')  e^{i(\mathbf k, \mathbf y)}\delta^{(2)} (\mathbf p_2 - \mathbf p_1 - \mathbf k) e^{i \alpha \tau} \delta (\omega_2 - \omega_1 -\alpha) \theta(\omega_2 - \omega_1) \ .
\end{gather} 
Moreover, one can take the integrals with the use of the $\delta$-function and get the following expression:
\begin{gather}
\mathbf{A}= 
- \int_{\mathbb R_+^3}  d\omega_1 d\omega_2 d\rho' \int d^2 \mathbf{p}_1 d^2 \mathbf p_2   \  [\mathbf{p}_1 + \mathbf{p}_2] [a^*_{\omega_1, \mathbf {p}_1}a_{\omega_2, \mathbf {p}_2} - a_{\omega_1, \mathbf {p}_1}a^*_{\omega_2, \mathbf {p}_2}] \frac{2 i  \rho'\sh \pi \alpha}{\pi |\mathbf p_2 -\mathbf p_1| } \times \nonumber \\
\times  K_{i \omega_1} ( \tilde p_1  \rho')  K_{i \omega_2} ( \tilde p_2  \rho')  K_{i(\omega_2 - \omega_1)} (|\mathbf p_2 -\mathbf p_1| \rho)K_{i(\omega_2 - \omega_1)} (|\mathbf p_2 -\mathbf p_1|\rho')  e^{i(\mathbf p_2 -\mathbf p_1, \mathbf y)} e^{i(\omega_2 - \omega_1)\tau} \theta(\omega_2 - \omega_1)  \ . 
\end{gather}
Finally, we can express the above result in terms of the differential operator's harmonics:
\begin{gather}
\mathbf A= -\int_{\mathbb R_+^2}  d\omega_1 d\omega_2 \int d^2 \mathbf{p}_1 d^2 \mathbf p_2 \ [\mathbf{p}_1 + \mathbf{p}_2] [a^*_{\omega_1, \mathbf {p}_1}a_{\omega_2, \mathbf {p}_2} - a_{\omega_1, \mathbf {p}_1}a^*_{\omega_2, \mathbf {p}_2}] \frac{2 i  \sh \pi \alpha}{\pi |\mathbf p_2 -\mathbf p_1| }  \gamma (x| \omega_2 - \omega_1, \omega_2 - \omega_1, \mathbf p_2 -\mathbf p_1)\times \nonumber \\
\times \theta(\omega_2 - \omega_1)\int_{\mathbb R_+} d \rho' \rho'  K_{i \omega_1} ( \tilde p_1  \rho')  K_{i \omega_2} ( \tilde p_2  \rho')   K_{i(\omega_2 - \omega_1)} (|\mathbf p_2 -\mathbf p_1|\rho') \ .
\end{gather}
So the vector field harmonics are radiated with the transversal momentum and energy equal to the corresponding gap between the scalar states.

Now we should conduct similar calculations for the zero component of the vector field. We start with:
\begin{gather}
A_0 = \int_{\mathbb R_+} d^+ \rho'\int d \tau' d^2 \mathbf y' G_0(x|x')  j_0(x') = \nonumber \\
=  -\int_{\mathbb R_+^4}  d\omega_1 d\omega_2 d\rho' \frac{d \alpha}{\pi^2} \int d^2 \mathbf{p}_1 d^2 \mathbf p_2 \frac{ d^2\mathbf{k}}{(2\pi)^2} d \tau'  d^2 \mathbf y' \ e [\omega_1 + \omega_2] a^*_{\omega_1, \mathbf {p}_1}a_{\omega_2, \mathbf {p}_2} \frac{i \alpha \rho\sh \pi \alpha}{|\mathbf k| \sqrt{1+\alpha^2} } \theta(\tau -\tau')  K_{i \omega_1} ( \tilde p_1  \rho')  K_{i \omega_2} ( \tilde p_2  \rho') \times \nonumber \\
\times K_{i\alpha} (|\mathbf k| \rho)K_{i\alpha} (|\mathbf k| \rho')  e^{i(\mathbf k, \mathbf y)} e^{i(\mathbf p_2 - \mathbf p_1 - \mathbf k, \mathbf y')}\left[ e^{i\sqrt{1+\alpha^2}\tau }e^{i(\omega_2 - \omega_1 - \sqrt{1+\alpha^2})\tau'} - e^{-i\sqrt{1+\alpha^2}\tau }e^{i(\omega_2 - \omega_1 + \sqrt{1+\alpha^2})\tau'} \right] \ .
\end{gather}
The integral over $d^2 \mathbf y'$ will lead to the same $\delta$-function as previously, but after we integrate over $d\tau'$ and take the limit we get:
\begin{gather}
\lim_{T \to \infty} \int_{-T}^T \frac{d \tau'}{2 \pi} e(\tau') \theta (\tau - \tau')\left[ e^{i\sqrt{1+\alpha^2}\tau }e^{i(\omega_2 - \omega_1 - \sqrt{1+\alpha^2})\tau'} - e^{-i\sqrt{1+\alpha^2}\tau }e^{i(\omega_2 - \omega_1 + \sqrt{1+\alpha^2})\tau'} \right]= \nonumber \\ = e^{i  \sqrt{1+\alpha^2} \tau} \delta (\omega_2 - \omega_1 - \sqrt{1+\alpha^2}) - e^{-i  \sqrt{1+\alpha^2} \tau} \delta (\omega_1 - \omega_2 -  \sqrt{1+\alpha^2}) \ .
\end{gather}
$\delta$-functions can be rewritten as follows:
\begin{gather}
\delta (\omega_2 - \omega_1 - \sqrt{1+\alpha^2}) = \frac{\sqrt{1 + \alpha^2}}{\alpha} [\delta(\alpha - \sqrt{(\omega_2-\omega_1)^2 - 1}) + \delta(\alpha + \sqrt{(\omega_2-\omega_1)^2 - 1})] \theta (\omega_2 - \omega_1 - 1) \ , \nonumber \\
\delta (\omega_1 - \omega_2 - \sqrt{1+\alpha^2}) = \frac{\sqrt{1 + \alpha^2}}{\alpha} [\delta(\alpha - \sqrt{(\omega_2-\omega_1)^2 - 1}) + \delta(\alpha + \sqrt{(\omega_2-\omega_1)^2 - 1})] \theta (\omega_1 - \omega_2- 1) \ . 
\end{gather}
But as we know under the integral $\alpha \in \mathbb R_+$, so after substituting this into (99) and making the change between $\omega_1$ and $\omega_2$ in the second summand, as previously, we end up with:
\begin{gather}
A_0 = -\int_{\mathbb R_+^4}  d\omega_1 d\omega_2 d\rho' \frac{d \alpha}{\pi} \int d^2 \mathbf{p}_1 d^2 \mathbf p_2 \ d^2\mathbf{k}  \  [\omega_1 + \omega_2] [a^*_{\omega_1, \mathbf {p}_1}a_{\omega_2, \mathbf {p}_2} - a_{\omega_1, \mathbf {p}_1}a^*_{\omega_2, \mathbf {p}_2}] \frac{2 i  \rho\sh \pi \alpha}{|\mathbf k| }   K_{i \omega_1} ( \tilde p_1  \rho')  K_{i \omega_2} ( \tilde p_2  \rho')\times \nonumber \\
\times K_{i\alpha} (|\mathbf k| \rho)K_{i\alpha} (|\mathbf k| \rho')  e^{i(\mathbf k, \mathbf y)}\delta^{(2)} (\mathbf p_2 - \mathbf p_1 - \mathbf k) e^{i \sqrt{1+\alpha^2} \tau} \delta(\alpha - \sqrt{(\omega_2-\omega_1)^2 - 1})\theta(\omega_2 - \omega_1-1)  \ .
\end{gather}
After the integration with the use of the $\delta$-functions we obtain:
\begin{gather}
A_0
= - \int_{\mathbb R_+^3}  d\omega_1 d\omega_2 d\rho' \int d^2 \mathbf{p}_1 d^2 \mathbf p_2  d \  [\omega_1 + \omega_2] [a^*_{\omega_1, \mathbf {p}_1}a_{\omega_2, \mathbf {p}_2} - a_{\omega_1, \mathbf {p}_1}a^*_{\omega_2, \mathbf {p}_2}] \frac{2 i  \rho\sh \pi \alpha}{\pi |\mathbf p_2 - \mathbf p_1| } 
  K_{i \omega_1} ( \tilde p_1  \rho')  K_{i \omega_2} ( \tilde p_2  \rho') \times \nonumber  \\
\times K_{i\sqrt{(\omega_2-\omega_1)^2 - 1}} (|\mathbf p_2 - \mathbf p_1| \rho)K_{i\sqrt{(\omega_2-\omega_1)^2 - 1}} (|\mathbf p_2 - \mathbf p_1| \rho')  e^{i(\mathbf p_2 - \mathbf p_1, \mathbf y)} e^{i (\omega_2 - \omega_1) \tau} \theta(\omega_2 - \omega_1-1) \ .
\end{gather}
Expressing this in  terms of the corresponding harmonics, we get:
\begin{gather}
A_ 0 = - \int_{ R_+^2}  d\omega_1 d\omega_2  \int d^2 \mathbf{p}_1 d^2 \mathbf p_2  [a^*_{\omega_1, \mathbf {p}_1}a_{\omega_2, \mathbf {p}_2} - a_{\omega_1, \mathbf {p}_1}a^*_{\omega_2, \mathbf {p}_2}] \frac{2 i \sh \pi \alpha}{\pi |\mathbf p_2 - \mathbf p_1|^2} \psi(x|\sqrt{(\omega_2-\omega_1)^2 - 1}, \omega_2 - \omega_1,\mathbf p_2 - \mathbf p_1  ) \times \nonumber  \\
\times [\omega_1 + \omega_2] \theta(\omega_2 - \omega_1 - 1) \int_{\mathbb R_+}  d \rho' K_{i \omega_1} ( \tilde p_1  \rho')  K_{i \omega_2} ( \tilde p_2  \rho') K_{i\sqrt{(\omega_2-\omega_1)^2 - 1}} (|\mathbf p_2 - \mathbf p_1| \rho') \ .
\end{gather}
So the harmonics with similar properties as in the transversal case are radiated, but only if the gap between $\omega_2$ and $\omega_1$ is bigger than\footnotemark[17] 1. The obtained expressions for $A_0$ and $\mathbf A$ do not vanish and lead to non-trivial electromagnetic fields $\mathbf E$ and $\mathbf B$.
\footnotetext[17]{It should be clarified that the frequencies of the harmonics in the Rindler frame, for example $\omega_1$, are dimensionless. This follows from the fact that the time coordinate itself is dimensionless due to the equality $\tau = \arcth (t/x)$.}
\subsection{Uniformly accelerated source in the Minkowski space-time}

We have already studied the case corresponding to a free motion in the Rindler frame. Now, we would like to inspect the same question from the viewpoint of the Minkowski space-time, i.e. to consider, as a source of radiation, the scalar field excitation which is moving with a uniform acceleration. 

In order to do so we let the vector field to be composed of two parts: a large uniform background and a small excitation produced by the scalar field i.e. $A_\mu = \tilde A_\mu + a_\mu $. We will substitute the background into the equations of motion for the scalar field, solve them and, afterwards, use the result to calculate the current which will in turn excite $a_\mu$.

We take $\tilde A_\mu$ to be of the form $(0,-Et,0,0)$, which corresponds to the uniform background electric field directed along the $x$-axis.
Now, we need to find the scalar field which is given by the following equation (we drop $a_\mu$ in the expressions because it is much smaller than $\tilde A_\mu$):
\begin{equation}
[m^2 + D[\tilde A]_\mu D[\tilde A]^\mu] \phi = 0 \ ,
\end{equation}
It can be rewritten as:
\begin{equation}
[m^2 + \alpha^2t^2 + \Box -2i\alpha t \pd_1] \phi = 0 ,
\end{equation}
with $\alpha = eE$.
Now, we shall choose the harmonics to be of the form $\phi = f_{\mathbf k}(t)e^{i(\mathbf k, \mathbf x)}$. In this case $f_\mathbf{k}(t)$ should satisfy the following equation:\footnote[18]{In this section bold letters stand for three-vectors	with the exception of $\mathbf k_{\perp}$ which stands for the transverse two-vector.}
\begin{equation}
[m^2 + \alpha^2 t^2 + \pd_t^2 + \mathbf k^2 + 2\alpha t k_1]f_{\mathbf k}(t) = 0
\end{equation}
We would like to find the function $f_{\mathbf k}(t) = f_{\mathbf k}(x^0)$ explicitly. After the substitution of $\vartheta= \sqrt{\alpha} t + \frac{k_1}{\sqrt{\alpha}}$ and $a =\frac{m^2 +\mathbf k^2_\perp}{\alpha} $ ($\mathbf k_\perp = (k_2, k_3)$) into the (107), we obtain:
\begin{equation}
f_{\mathbf k}'' +\left (a + \vartheta^2\right)f_{\mathbf k} = 0 \ .
\end{equation}
The solution of this equation can be represented as a linear combination of $e^{\frac{-i\vartheta^2}{2}} {}_1F_1\left(\frac{ai+1}{4}, \frac{1}{2}, i\vartheta^2\right)$  which behaves like a cosine for small $\vartheta$ , and  $ \vartheta e^{\frac{-i\vartheta^2}{2}} {}_1F_1\left(\frac{ai+3}{4}, \frac{3}{2}, i\vartheta^2\right)$  which behaves like a sine. For $\vartheta \ll a$ the equation for $f_{\mathbf k}$ tends to the harmonic oscillator equation with the frequency $\sqrt{a}$, so we would like our function to behave like an $e^{i\sqrt{a}\vartheta}$ for small $\vartheta$. In other words, at short times (high energies) the harmonics that we would like to consider should behave as plane waves. This could be satisfied by the following combination:\footnote[19]{It is probably worth stressing that we obtain a non-trivial result for the vector field independently of the choice of the time dependent part of the harmonic.}
\begin{equation}
f_{\mathbf k} = e^{\frac{-i\vartheta^2}{2}} \left [{}_1F_1\left(\frac{ai+1}{4}, \frac{1}{2}, i\vartheta^2\right) + i \sqrt{a}\vartheta \ {}_1F_1\left(\frac{ai+3}{4}, \frac{3}{2}, i\vartheta^2\right) \right] \ .
\end{equation}
Then, we rewrite it in terms of $t$ and $\alpha$:
\begin{gather}
f_{\mathbf k} =  \exp\left[\frac{-i(\alpha t + k_1)^2}{2\alpha}\right] \left [{}_1F_1\left(\frac{im^2 +i\mathbf k^2_\perp+\alpha}{4\alpha}, \frac{1}{2}, \frac{i(\alpha t + k_1)^2}{\alpha}\right) \right.+ \\  \left.+  i  \frac{\sqrt{m^2 +\mathbf k^2_\perp}}{\alpha}(\alpha t + k_1) \ {}_1F_1\left(\frac{im^2 +i\mathbf k^2_\perp+3\alpha}{4\alpha}, \frac{3}{2}, \frac{i(\alpha t + k_1)^2}{\alpha}\right)  \right] \ .
\end{gather}
Thus, we choose $\phi$ to be of the form:
\begin{equation}
\phi = \int d^3\mathbf p a_\mathbf p f_{\mathbf p} (x^0) e^{i (\mathbf x, \mathbf p)} \ .
\end{equation}
For this choice of $\phi$ the 3-vector part of the current is expressed as follows:
\begin{equation}
\mathbf j = \int d^3 \mathbf p_1 d^3\mathbf p_2  a^*_{\mathbf p_1} a_{\mathbf p_2}(\mathbf p_1 + \mathbf p_2) e^{i(p_2-p_1)} f^*_{\mathbf p_1}(x^0) f_{\mathbf p_2}(x^0) \ .
\end{equation}
We substitute this current into (63) and integrate over $d^3\mathbf y d^3 \mathbf k$ in order to obtain the momentum conservation law, like in the case of the plane waves. Afterwards, we get the following expression:
\begin{equation}
\mathbf  a = - \int d^3 \mathbf p_1 d^3\mathbf p_2  d y^0 \frac{i e(y^0)}{2|\mathbf p_2 - \mathbf p_1|} e^{i(\mathbf p_2 - \mathbf p_1, \mathbf x)} \theta (x^0 - y^0)a^*_{\mathbf p_1} a_{\mathbf p_2}(\mathbf p_1 + \mathbf p_2) [e^{i|\mathbf p_2- \mathbf p_1|(x^0-y^0)}- c.c ] f^*_{\mathbf p_1}(y^0) f_{\mathbf p_2}(y^0) \ .
\end{equation}
By substituting the explicit form of $e(y^0)$  and taking the limit as before and rearranging the summands we obtain the following result for the vector field:
 \begin{gather}
 \mathbf  a =  -\int d^3 \mathbf{p}_1 d^3 \mathbf{p}_2 \frac{i e(\mathbf p_1 + \mathbf p_2)^*_{\mathbf p_1} a_{\mathbf p_2}}{2|\mathbf p_2 - \mathbf p_1|} \left [e^{i|\mathbf p_2- \mathbf p_1|x^0+ i(\mathbf p_2 - \mathbf p_1, \mathbf x)} \cdot \int dy^0 e^{-i|\mathbf p_2- \mathbf p_1|y^0}f^*_{\mathbf p_1}(y^0) f_{\mathbf p_2}(y^0) \ - \right. \nonumber \\ \left.
 - e^{-i|\mathbf p_2- \mathbf p_1|x^0+ i(\mathbf p_2 - \mathbf p_1, \mathbf x)} \cdot \int dy^0 e^{i|\mathbf p_2- \mathbf p_1|y^0}f^*_{\mathbf p_1}(y^0) f_{\mathbf p_2}(y^0)  \right] \ .
 \end{gather}
Performing the same manipulations for the zero component of the current we get:
 \begin{equation}
j_0 = -i e(y^0) \int d^3 \mathbf p_1 d^3\mathbf p_2  a^*_{\mathbf p_1} a_{\mathbf p_2} e^{i(p_2-p_1)} [f^*_{\mathbf p_1}(x^0) f'_{\mathbf p_2}(x^0) - f'^*_{\mathbf p_1}(x^0) f_{\mathbf p_2}(x^0)] \ ,
\end{equation}
hence,
\begin{equation}
  a_0 = -\int d^3 \mathbf p_1 d^3\mathbf p_2  d y^0 \frac{e(y^0)}{2|\mathbf p_2 - \mathbf p_1|} e^{i(\mathbf p_2 - \mathbf p_1, \mathbf x)} \theta (x^0 - y^0)a^*_{\mathbf p_1} a_{\mathbf p_2} [e^{i|\mathbf p_2- \mathbf p_1|(x^0-y^0)}- c.c ] [f^*_{\mathbf p_1}(x^0) f'_{\mathbf p_2}(x^0) - f'^*_{\mathbf p_1}(x^0)  	f_{\mathbf p_2}(x^0)] \ ,
 \end{equation}
and, finally,
  \begin{gather}
  a_0 = -\int d^3 \mathbf p_1 d^3\mathbf p_2 \frac{e a^*_{\mathbf p_1} a_{\mathbf p_2}}{2|\mathbf p_2 - \mathbf p_1|} \left [e^{i|\mathbf p_2- \mathbf p_1|x^0+ i(\mathbf p_2 - \mathbf p_1, \mathbf x)} \cdot \int dy^0 e^{-i|\mathbf p_2- \mathbf p_1|y^0}[f'_{\mathbf p_2}(y^0) f^*_{\mathbf p_1}(y^0) - f'^*_{\mathbf p_1}(y^0) f_{\mathbf p_2}(y^0)]  -\right. \nonumber \\  \left. -
 e^{-i|\mathbf p_2- \mathbf p_1|x^0+ i(\mathbf p_2 - \mathbf p_1, \mathbf x)} \cdot \int dy^0 e^{i|\mathbf p_2- \mathbf p_1|y^0}[f'_{\mathbf p_2}(y^0) f^*_{\mathbf p_1}(y^0) - f'^*_{\mathbf p_1}(y^0) f_{\mathbf p_2}(y^0)] \right] \ .
 \end{gather}
Thus, we have obtained the sum of two vector harmonics with the momentum equal to the gap between the momentum of the scalar states. The expressions for $a_0$ and $\mathbf a$ do not vanish and lead to non-trivial $\mathbf E$ and $\mathbf B$ as in the previous subsection.

\subsection{Discussion}

From our considerations, it follows that in the flat space-time a free scalar field does not produce an excitation of the vector harmonics, but uniformly accelerated scalar harmonics create a non-zero vector field. This observation corresponds to the fact that free particles in the flat space-time do not radiate, but uniformly accelerated ones do. 
For the general situation in the Rindler frame the resulting vector field is non-zero. However, this effect can be attributed to the transversal part of the currents. Indeed, if one restricts his consideration to the scalar harmonics with $\mathbf p = 0$ one will get that $\gamma (x| \omega_2 - \omega_1, \omega_2 - \omega_1, 0) = K_{i(\omega_2-\omega_1)}(0\cdot \rho)e^{i(\omega_2-\omega_1)t}$. The latter expression does not depend on spatial coordinates and leads to a vanishing  $a_0$ and $\mathbf a$ at the future infinity. In other words, the radiation in the co-moving Rindler frame is not possible only if the value of the transversal momentum stays constant. This result nicely matches with the point-like source case. Indeed, in the first part of this work we have considered a point-like stationary particle (transverse momentum is zero) and the radiation was absent. A particle moving uniformly in the transverse direction in the Rindler coordinates has a complicated world line in the Minkowski space-time, so it is not trivial to find the fields it generates. However, this correspondence can be used to predict that the transversely moving point-like particle radiates even in the Rindler frame.
 
\section{Acknowledgements}
I would like to thank E.T. Akhmedov for proposing this problem to me and for the valuable discussions on this topic. Moreover, I am most grateful to E.~T.~Akhmedov and O.~R.~Galeev for proof reading this text. Besides, I would like to express my gratitude to the Albert Einstein Institute for their hospitality during the final stage of the work on this project.

\end{document}